\documentclass[12pt]{article}
\usepackage{epsfig,amsfonts,amssymb}
\usepackage{hyperref}

\usepackage{cite}
\usepackage{bm}
\usepackage[table]{xcolor}

\usepackage{amsmath}
\usepackage{bigints}

\usepackage[mathscr]{euscript}
 \let\mathscr\relax
\usepackage[scr]{rsfso}
\newcommand{\powerset}{\raisebox{.15\baselineskip}{\Large\ensuremath{\wp}}}

\begin{document}

\baselineskip 24pt

\begin{center}

{\Large \bf Analyticity of Off-shell Green's Functions in Superstring Field Theory}

\end{center}

\vskip .6cm
\medskip

\vspace*{4.0ex}

\baselineskip=18pt

\centerline{\large \rm Ritabrata Bhattacharya$^*$ and Ratul Mahanta$^\dagger$}

\vspace*{4.0ex}

\centerline{\large $^*$Chennai Mathematical Institute,}
\centerline{\large H1, SIPCOT IT Park, Siruseri, Kelambakkam 603103, India}

\vspace*{1.0ex}
\centerline{\large $^\dagger$Harish-Chandra Research Institute, HBNI,}
\centerline{\large Chhatnag Road, Jhunsi, Allahabad 211019, India}

\begin{NoHyper}
  \let\thefootnote\relax\footnotetext{electronic address: {\tt {ritabratab@cmi.ac.in, ratulmahanta@hri.res.in} }}
\end{NoHyper}

\vspace*{5.0ex}

\thispagestyle{empty}
\centerline{\bf Abstract} \bigskip

We consider the off-shell momentum space Green's functions in closed superstring field theory. Recently in \cite{LES2019}, the off-shell Green's functions --- after explicitly removing contributions of massless states --- have been shown to be analytic on a domain (to be called the LES domain) in complex external momenta variables. Analyticity of off-shell Green's functions in local QFTs without massless states in the primitive domain is a well-known result. Using complex Lorentz transformations and Bochner's theorem allow to extend the LES domain to a larger subset of the primitive domain. For the 2-, 3- and 4-point functions, the full primitive domain is recovered. For the 5-point function, we are not able to obtain the full primitive domain analytically, only a large part of it is recovered. While this problem arises also for higher-point functions, it is expected to be only a technical issue.

\vfill \eject

\baselineskip 18pt

\setcounter{page}{1}
\tableofcontents

\section{Introduction}
\label{intro}

Closed superstring field theory (SFT) which is designed to reproduce perturbative amplitudes of superstring theory is a quantum field theory with countable infinite number of fields and non-local interaction vertices, whose action is directly written in momentum space (for a detailed review, see \cite{LEKSV2017}). It contains massless states. Off-shell momentum space amputated $n$-point Green's function\footnote{Hereafter Green's functions will always refer to momentum space amputated Green's functions.} in SFT is defined by usual momentum space Feynman rules \cite{LEKSV2017,PS2016}. It can be computed by summing over connected Feynman diagrams with $n$ amputated external legs carrying ingoing $D$-momenta $p_1,\dots,p_n$.

We consider the off-shell $n$-point Green's function $G(p_1,\dots,p_n)$ in SFT after explicitly removing contributions coming from one or more massless internal propagators (for more details, see \cite{LES2019}). That is, the relevant part of the perturbative expansion of the off-shell $n$-point Green's function keeps only those Feynman diagrams which do not contain any internal line corresponding to a massless particle. We call this part of the off-shell Green's function as the infrared safe part\footnote{Hereafter off-shell Green's functions in SFT will refer to this part of respective off-shell Green's functions in SFT, if not explicitly stated. This part of the off-shell Green's functions --- when all the external particles are massless --- precisely gives the vertices of the Wilsonian effective field theory of massless fields, obtained by integrating out the massive fields in superstring theory \cite{LEKSV2017,Sen2017}.}. In \cite{LES2019} de Lacroix, Erbin and Sen (LES) showed that the infrared safe part of the off-shell $n$-point Green's function $G(p_1,\dots,p_n)$ in SFT as a function of $(n-1)D$ complex variables (taking into account the momentum conservation $\sum_{a=1}^np_a=0$ for external momenta) is analytic on a domain which we call as the LES domain. At the heart of this result, it has been proven that each of the relevant Feynman diagrams $\text{F}(p_1,\dots,p_n)$ has an integral representation in terms of loop integrals as presented below, whenever the external momenta lie on the LES domain.
\begin{equation}
\label{eq:feyn}
\begin{aligned}
  \text{F}(p_1,\dots,p_n)
  &=\ \bigintssss \prod_{r=1}^L\frac{d^Dk_r}{(2\pi)^D}\frac{f(k_1,\dots,k_L;p_1,\dots,p_n)}{\prod_{s=1}^\mathscr{I}\ \left(\big(\ell_s(\{k_r\};\{p_a\})\big)^2 + m_s^2\right)}\ .
\end{aligned}
\end{equation}
The above analytic function $\text{F}$ multiplied by a factor of $(2\pi)^D\delta^{(D)}(p_1+\cdots+p_n)$ gives the usual Feynman diagram. Equation (\ref{eq:feyn}) represents an $n$-legged $L$-loop graph with $\mathscr{I}$ internal lines, where $k_r$ is a loop momentum, $\ell_s$ is the momentum of the internal line with mass $m_s\ (\neq0)$\footnote{In our notation, $\ell_s^2=-(\ell^0_s)^2+\sum_{i=1}^{D-1}(\ell^i_s)^2$ for each $s=1,\dots,\mathscr{I}$.}, and $f$ is a regular function whenever its arguments take finite complex values. The function $f$ contains the product of the vertex factors associated with the vertices of the graph, as well as possible momentum dependence from the numerators of the internal propagators. The momentum $\ell_s$ of an internal line is usually a linear combination of the loop momenta and the external momenta. Due to certain non-local properties of the vertices in SFT, the graph is manifestly UV finite as long as for each $r$, $k_r^0$ integration contour ends at $\pm i\infty$, and $k_r^i,\ i=1,\dots,(D-1)$ integration contours end at $\pm\infty$. The prescription for the loop integration contours has been given as follows \cite{PS2016}. At origin, i.e. $p_a=0\ \forall a=1,\dots,n$ each loop energy integral is to be taken along the imaginary axis from $-i\infty$ to $i\infty$ and each spatial component of loop momenta is to be taken along the real axis from $-\infty$ to $\infty$. With this, $\text{F(0,\dots,0)}$ has been shown to be finite as all the poles of the integrand in any complex $k_r^\mu$ plane are at finite distance away from the loop integration contour. As we vary the external momenta from the origin to other complex values if some of such poles approach the $k_r^\mu$ contour, the contour has to be bent away from those poles keeping its ends fixed at $\pm i\infty$ for loop energies and $\pm\infty$ for loop momenta. It has been shown that there exists a path inside the LES domain connecting the origin to any other point $p\equiv(p_1,\dots,p_n)$ of the LES domain such that when we vary external momenta from origin to that point $p$ along that path, the loop integration contours in any graph can be deformed away avoiding poles of the integrand which approach them \cite{LES2019}. Hence the integral representation (\ref{eq:feyn}) for $\text{F}(p_1,\dots,p_n)$ when the external momenta lie on the LES domain is well defined where the poles of the integrand are at finite distance away from the (deformed)loop integration contours.

On the other hand in local quantum field theories {\it without} massless particles, the off-shell $n$-point Green's function $G(p_1,\dots,p_n)$ as a function of $(n-1)D$ complex variables (external ingoing $D$-momenta $p_1,\dots,p_n$ subject to momentum conservation) is known to be analytic on a domain called the primitive domain \cite{Bogolyubov, Steinmann1960, Ruelle1961, AB1960, Araki1961}. This result follows from causality constraints on the position space Green's functions in a local QFT and representing the momentum space Green's functions as Fourier transforms of the position space correlators\footnote{But the lack of a position space description of closed superstring field theory forces us to work directly in the momentum space.}. The primitive domain contains the LES domain as a proper subset. In several complex variables, the analyticity domain cannot be arbitrary. For example, the shape of the primitive domain allows the actual domain of holomorphy of $G(p_1,\dots,p_n)$ to be larger than itself (e.g. \cite{BMS1961,BEG1964}). This can be used to prove various analyticity properties \cite{BEG1965, Bros1986, Pavlov1978, MP1978, LMMPS1979, Bros1980, MPPS1982, MPPS1984} of the S-matrix of QFTs, since the S-matrix is defined as the on-shell connected $n$-point Green's function for $n\ge 3$. These properties are to be read as the artifact of the shape of the primitive domain, irrespective of the functional form of $G(p_1,\dots,p_n)$ which is defined on it. In particular, the derivations \cite{BMS1961, BEG1965, Bros1986}\footnote{For example, \cite{BMS1961} recovered the JLD domain, \cite{BEG1965} proved the crossing symmetry of the $2\to2$ scattering amplitudes.} of certain analyticity properties \cite{BEG1965, Bros1986, JL1957, Dyson1958} of the S-matrix use the information of only the LES domain as a subregion of the primitive domain.

From \cite{LES2019} we already know that the infrared safe part of the off-shell $n$-point Green's function in SFT is analytic on the LES domain. Hence \cite{LES2019} basically established that in superstring theory any possible departure from those analyticity properties of the full S-matrix that rely only on the LES domain is entirely due to the presence of massless states. This is also true for local QFTs that have massless states. Thus, with respect to the aforenamed analytic properties \cite{BEG1965, Bros1986, JL1957, Dyson1958}, the S-matrix of the superstring theory displays similar behaviour to that of a standard local QFT with massless particles. In local QFTs {\it without} massless states, analyticity properties  \cite{Pavlov1978, MP1978, LMMPS1979, Bros1980, MPPS1982, MPPS1984} of the S-matrix rely on the properties of the primitive domain that are not restricted to its LES subregion. We know that local QFTs {\it with} massless states could possibly deviate from these properties for their full S-matrix. However, any such departures are entirely due to the presence of massless states, i.e. the infrared safe parts of respective amplitudes in local QFTs with massless states must satisfy all these analyticity properties. At this stage, it is natural to ask that whether the infrared safe part of the S-matrix of the superstring theory (despite having non-local vertices) satisfy the last-mentioned analyticity properties, or not. They satisfy them, only if the relevant part of the off-shell $n$-point Green's function in SFT can be shown to be analytic on the full primitive domain extending the LES domain.

In this paper, we aim to generalize the result of \cite{LES2019} by showing that the infrared safe part of the off-shell $n$-point Green's function in SFT is analytic on a larger domain than the LES domain. As will be reviewed in section \ref{LESD}, the analyticity property of the off-shell Green's functions in SFT is invariant under the action of a $D$-dimensional complex Lorentz transformation on all the external momenta. Thus the off-shell Green's functions in SFT are also analytic at points that are obtained by the action of such transformations on points in the LES domain \cite{LES2019}. We consider LES domain adjoining these new points. The primitive domain essentially contains the union of a certain family of convex tube domains. We call such tubes as the primitive tubes. Within each such primitive tube, we identify a connected tube which is also contained in the LES domain and its shape allows us to carry out a holomorphic extension to its convex hull inside the corresponding primitive tube, due to a classic theorem by Bochner \cite{Bochner1937}. The domain thus found may still be smaller. We explicitly work out the cases of the three-, four- and five-point functions to determine whether such convex hulls fully obtain respective primitive tubes, or not. The extension to the primitive domain is trivial for the two-point function. In the case of three-point function, indeed such extensions yield all the 6 possible primitive tubes. Also for the four-point function, by such extensions all the 32 possible primitive tubes are obtained. Whereas for the five-point function, out of 370 possible primitive tubes, for 350 of them we are able to show that such extensions obtain each of them fully. The technique that we employ for aforesaid checks seems difficult to implement analytically for the remaining 20 primitive tubes whose shapes are complicated. This technical difficulty is a feature of the higher point functions as well. However in any case, our work establishes that based on a geometric consideration only, the LES domain is holomorphically extended inside all the primitive tubes. This extension does not depend on the details of the Green's functions. Thus with respect to all the analyticity properties of the S-matrix which can be obtained from this extended domain\footnote{Note that for the 2-, 3- and 4-point functions, the extended domain is equal to the primitive domain.}, superstring theory behaves like a standard local QFT that has massless states.

We organize the paper as follows. In section \ref{Rev}, we briefly review certain properties of the primitive domain and the LES domain which are useful for our purpose. In section \ref{HExtnLESD}, we start with the general scheme to extend the LES domain holomorphically. We apply this scheme to the case of the three-point function in subsection \ref{3ptfun}. In subsection \ref{4ptfun} (and appendix \ref{4ptTbases}) we deal with the case of the four-point function, and in subsection \ref{5ptfun} the case of five-point function. We discuss certain real limits within each of the primitive tubes in section \ref{limTlamb}.

\section{Review}
\label{Rev}

In this section we review certain properties of the two domains, namely the primitive domain and the LES domain which will be useful for our analysis. Both the domains are domains in the complex manifold $\mathbb{C}^{(n-1)D}$ given by $p_1+\cdots+p_n=0$. The origin of $\mathbb{C}^{(n-1)D}$ which is given by $p_a=0\ \forall a=1,\dots,n$ will be denoted by $O$. We count $p_a$ as positive if ingoing, negative otherwise. We shall use Minkowski metric with mostly plus signature.

\subsection{The primitive domain}
\label{PD}

The primitive domain $\mathcal{D}$ is given by
\begin{equation}
\label{eq:PD}
\begin{aligned}
  \mathcal{D}&=&\bigg\{&p\equiv(p_1,\dots,p_n):\ {\sum}_{a=1}^n p_a=0\ \text{and for each}\ I\in\powerset^*(X)\\
    &&&\text{either,}\ \text{Im }P_I\neq0,\ (\text{Im }P_I)^2\leq0\quad \text{or,}\ \text{Im }P_I=0,\ -P_I^2<M_I^2\bigg\}\ ,
\end{aligned}
\end{equation}
where $X=\{1,\dots,n\}$ is the set of first $n$ natural numbers. $\powerset^*(X)=\{I\subsetneq X,\ \text{except}\ \emptyset\}$ is the collection of all non-empty proper subsets of $X$. $P_I$ is defined to be equal to $\sum_{a\in I}p_a$. $M_I$ is the threshold of production of any (multi-particle) state in a channel containing the external states in the set $\{p_a,\ a\in I\}$, i.e. $M_I$ is the invariant threshold mass for producing any set of intermediate states in the collision of particles carrying total momentum $P_I$.

Clearly, $O\in\mathcal{D}$. Primitive domain is star-shaped with respect to $O$, i.e. the straight line segment connecting $O$ and any point $p\in\mathcal{D}$ which is given by $\big\{tp:\ t\in[0,1]\big\}$ lie entirely inside $\mathcal{D}$. Hence the primitive domain is path-connected as any two points $p^{(1)},\ p^{(2)}\in\mathcal{D}$ can be connected via the straight line segments $p^{(1)}O$ and $Op^{(2)}$. Furthermore primitive domain is simply connected, i.e. any closed curve within $\mathcal{D}$ can be continuously shrunk to the point $O$, which is a property of a star-shaped domain.

Primitive domain essentially contains the union of a family of mutually disjoint tube\footnote{A subset of $\mathbb{C}^m$ is called a tube if it is equal to $\mathbb{R}^m+iA$ for some subset $A$ of $\mathbb{R}^m$ where $m$ is a given natural number. $A$ is called the base of the tube.} domains denoted by $\{\mathcal{T}_\lambda,\ \lambda\in\Lambda^{(n)}\}$ \cite{AB1960, BEG1964, Lassalle1974, BL1975}. Any member $\mathcal{T}_\lambda$ of this family will be called a primitive tube. To describe this family of tubes the following definitions are needed. We consider the space $\mathbb{R}^{n-1}$ of $n$ real variables $s_1,\dots,s_n$ linked by the relation $s_1+\cdots+s_n=0$. We define $S_I=\sum_{a\in I}s_a$ for each $I\in\powerset^*(X)$. The family of planes $\{S_I=0,\ I\in\powerset^*(X)\}$ (where the planes $S_I=0$ and $S_{X\setminus I}=0$ are identical) divides the above space $\mathbb{R}^{n-1}$ into open convex cones\footnote{\label{fn:cone}A subset $A$ of $\mathbb{R}^m$ is called a cone if any point $p\in A\implies\alpha p\in A\ \forall\alpha>0$.} with common apex at the origin. Any such cone will be called a cell, $\gamma_\lambda$. Within a cell each $S_I$ is of definite sign $\lambda(I)$. Thus a cell can be written as
\begin{equation}
  \gamma_{_\lambda}=\bigg\{s\in\mathbb{R}^{n-1}:\ \lambda(I)S_I>0\quad \forall I\in\powerset^*(X)\bigg\}\ ,
\end{equation}
where $\lambda:\powerset^*(X)\to\{-1,1\}$ is a sign-valued map with the following properties.
\begin{equation}
\label{eq:lambprop}
\begin{aligned}
  &(\text{i})\ \forall I\in\powerset^*(X)\quad \lambda(I)=-\lambda(X\setminus I),\\
  &(\text{ii})\ \forall I,J\in\powerset^*(X)\ \text{with}\ I\cap J=\emptyset\ \text{and}\ \lambda(I)=\lambda(J)\quad \lambda(I\cup J)=\lambda(I)=\lambda(J)\ .
\end{aligned}
\end{equation}
The first property is compatible with $s_1+\cdots+s_n=0$ and the second property is compatible with $S_{I\cup J}=S_I+S_J$ whenever $I\cap J=\emptyset$. $\Lambda^{(n)}$ denotes the collection of all possible maps $\lambda$ satisfying above properties. Corresponding to each cell $\gamma_\lambda$, now we associate an open convex tube domain $\mathcal{T}_\lambda$ (primitive tube) given by\footnote{With $i=\sqrt{-1}$.}
\begin{equation}
\label{eq:primTlamb}
\begin{aligned}
  \mathcal{T}_\lambda=&\ \bigg\{p\equiv(p_1,\dots,p_n):\ {\sum}_{a=1}^n p_a=0,\ \lambda(I)\text{Im }P_I\in V^+\quad \forall I\in\powerset^*(X)\bigg\}\\
  =&\ \ \mathbb{R}^{(n-1)D}\ +\ i\mathcal{C}_\lambda\ ,\\
  \mathcal{C}_\lambda=&\ \bigg\{\text{Im }p\in\mathbb{R}^{(n-1)D}:\ \lambda(I)\text{Im }P_I\in V^+\quad \forall I\in\powerset^*(X)\bigg\}\ ,
\end{aligned}
\end{equation}
where $V^+$ is the open forward lightcone in $\mathbb{R}^D$. $\mathcal{C}_\lambda$ is the conical base of the tube $\mathcal{T}_\lambda$. Although the primitive domain is non-convex the primitive tubes $\mathcal{T}_\lambda$ are convex (see appendix \ref{convTlamb}). Hence each tube $\mathcal{T}_\lambda$ is path-connected as the entire straight line segment $p^{(1)}p^{(2)}$ connecting any two points $p^{(1)},\ p^{(2)}\in\mathcal{T}_\lambda$ is contained in the tube $\mathcal{T}_\lambda$.

\subsection{The LES domain}
\label{LESD}

The LES domain is given by
\begin{equation}
\label{eq:LESD}
\begin{aligned}
  \mathcal{D}'=\bigg\{p\equiv(p_1,\dots,p_n):\ \forall a\ \text{Im }p_a^\mu=0,\ \mu\neq0,1;\ {\sum}_{a=1}^n p_a=0\ \text{and for each}\ I\in\powerset^*(X)\\
  \text{either,}\ \text{Im }P_I\neq0,\ (\text{Im }P_I)^2\leq0\quad \text{or,}\ \text{Im }P_I=0,\ -P_I^2<M_I^2\bigg\}\ ,
\end{aligned}
\end{equation}
where all the $\text{Im }p_a$ thereby $\text{Im }P_I$ are allowed to lie only on the two dimensional Lorentzian plane\footnote{By a two dimensional Lorentzian plane we refer to any two dimensional plane in $\mathbb{R}^D$ which contains the $p^0$-axis.} $p^0-p^1$. Clearly $O\in\mathcal{D}'$ and the LES domain $\mathcal{D}'$ is contained in the primitive domain $\mathcal{D}$.

In \cite{LES2019} it has been argued that the domain of holomorphy of the $n$-point Green's function $G(p_1,\dots,p_n)$ in SFT is a connected region in $\mathbb{C}^{(n-1)D}$ containing the origin, and it is invariant under the action of Lorentz transformations $\tilde{\Lambda}$ with complex parameters, i.e. $\tilde{\Lambda}$ is any complex matrix satisfying $\tilde{\Lambda}^T\eta\tilde{\Lambda}=\eta$ for $\eta$ being the Minkowski metric in $\mathbb{R}^D$. We call the set of such matrices the complex Lorentz group, $\mathfrak{L}$. In general, the action of a complex Lorentz transformation $\tilde{\Lambda}$ is defined on the complex manifold $\mathbb{C}^{(n-1)D}$ taking a point to another point of the same manifold given by
\begin{equation}
\label{eq:actioncompLT}
  (p_1,\dots,p_n)\longmapsto(\tilde{\Lambda}p_1,\dots,\tilde{\Lambda}p_n)\ ,
\end{equation}
which we abbreviate as $p\mapsto\tilde{\Lambda}p$. Note that the same $\tilde{\Lambda}$ acts on all $p_a$.

As as consequence, the result of \cite{LES2019} automatically generalizes to a larger domain than the LES domain $\mathcal{D}'$, i.e. $G(p)$ is analytic on the domain $\tilde{\mathcal{D}}'$ given by
\begin{equation}
\label{eq:complorentz}
  \tilde{\mathcal{D}}'=\bigg\{\tilde{\Lambda}p:\ p\in\mathcal{D}',\ \tilde{\Lambda}\in\mathfrak{L}\bigg\}\ .
\end{equation}
Clearly $\tilde{\mathcal{D}}'\supset\mathcal{D}'$, since $\mathfrak{L}$ contains the identity matrix. Hereafter we refer $\tilde{\mathcal{D}}'$ as the LES domain.

\section{Extension of the LES domain \texorpdfstring{$\tilde{\mathcal{D}}'$}{TEXT}}
\label{HExtnLESD}

We identify a family of tubes lying inside the primitive tube $\mathcal{T}_\lambda$ which is a convex tube domain (described by equation (\ref{eq:primTlamb}) in section \ref{PD}) such that any member of the family is also contained in the LES domain $\tilde{\mathcal{D}}'$ (described in section \ref{LESD}). Any member of this family is convex and it can be characterized by a set of $D-2$ angles $\vec{\theta}$ which specifies a two dimensional Lorentzian plane $p^0-p^{\vec{\theta}}$\footnote{Here the axis $p^{\vec{\theta}}$ lies in the subspace $\mathbb{R}^{D-1}$ of points $(p^1,\dots,p^{D-1})$. It can be specified by a point on the unit sphere $S^{D-2}$ in $\mathbb{R}^{D-1}$. With a set of $D-2$ angles $(\theta_1,\dots,\theta_{D-2})\equiv\vec{\theta}$ where $0\leq\theta_1,\dots,\theta_{D-3}\leq\pi$ and $0\leq\theta_{D-2}<2\pi$, the axis $p^{\vec{\theta}}$ can be explicitly written as
$$p^{\vec{\theta}}=(\cos\theta_1,\ \sin\theta_1\cos\theta_2,\ \sin\theta_1\sin\theta_2\cos\theta_3,\ \dots,\ \sin\theta_1\cdots\sin\theta_{D-3}\cos\theta_{D-2})\ .$$} with $\vec{\theta}=0$ specifying the $p^0-p^1$ plane. Hence we denote a member by $\mathcal{T}_\lambda^{\vec{\theta}}$. The convex tube $\mathcal{T}_\lambda^{\vec{\theta}}$ is given by
\begin{equation}
\label{eq:LESTlamb}
\begin{aligned}
  \mathcal{T}_\lambda^{\vec{\theta}}=&\ \bigg\{p\equiv(p_1,\dots,p_n):\ \forall a\ \text{Im }p_a\in\ p^0-p^{\vec{\theta}}\ \text{plane};\ {\sum}_{a=1}^np_a=0,\\
  &\quad\quad\quad\quad\quad\quad\quad\quad\quad\quad\quad\quad\text{and}\ \lambda(I)\text{Im }P_I\in V^+\quad \forall I\in\powerset^*(X)\bigg\}\\
  =&\ \ \mathbb{R}^{(n-1)D}\ +\ i\mathcal{C}_\lambda^{\vec{\theta}}\ ,\\
  \mathcal{C}_\lambda^{\vec{\theta}}=&\ \bigg\{(\text{Im }p_1,\dots,\text{Im }p_n)\ \text{on manifold}\ {\sum}_{a=1}^n\text{Im }p_a=0\ \text{such that}\\
    &\quad\quad\quad\forall a\ \text{Im }p_a\in\ p^0-p^{\vec{\theta}}\ \text{plane}\ \text{and}\ \lambda(I)\text{Im }P_I\in V^+\quad \forall I\in\powerset^*(X)\bigg\}\ ,
\end{aligned}
\end{equation}
where the base $\mathcal{C}_\lambda^{\vec{\theta}}$ is a subset of $\mathbb{R}^{(n-1)D}$. Any such tube $\mathcal{T}_\lambda^{\vec{\theta}}$ can be obtained by acting a real rotation on the tube $\mathcal{T}_\lambda^{\vec{\theta}=0}$.

Clearly $\bigcup_{\vec{\theta}}\mathcal{T}_\lambda^{\vec{\theta}}$ lies inside the primitive tube $\mathcal{T}_\lambda$ as well as it is contained in the LES domain $\tilde{\mathcal{D}}'$. Now $\bigcup_{\vec{\theta}}\mathcal{T}_\lambda^{\vec{\theta}}$ is a tube given by $\mathbb{R}^{(n-1)D}+i(\bigcup_{\vec{\theta}}\mathcal{C}_\lambda^{\vec{\theta}})$. Although each tube $\mathcal{T}_\lambda^{\vec{\theta}}$ is convex (see appendix \ref{convTlamb}) the tube $\bigcup_{\vec{\theta}}\mathcal{T}_\lambda^{\vec{\theta}}$ is non-convex\footnote{$\bigcup_{\vec{\theta}}\mathcal{T}_\lambda^{\vec{\theta}}$ is non-convex when $n>2$.} (see appendix \ref{nconvunionTlamb}). However the tube $\bigcup_{\vec{\theta}}\mathcal{T}_\lambda^{\vec{\theta}}$ is path-connected (see appendix \ref{pconnecunionTlamb}).

We apply Bochner's tube theorem \cite{Bochner1937,Rudin1971}\footnote{In \cite{Rudin1971} it has been stated as the `convex tube theorem' at the end of its third section. This version of the theorem is suitable for our purpose.} on the connected tube $(\bigcup_{\vec{\theta}}\mathcal{T}_\lambda^{\vec{\theta}})$\footnote{This tube can be thickened in order to make it open (see appendix \ref{thickunionTlamb}).}. The theorem states that any open connected tube $\mathbb{R}^m+iA$ has a holomorphic extension\footnote{By a holomorphic extension $\Omega'$ of a domain $\Omega$ in $\mathbb{C}^m$ we mean any larger domain $\Omega'$ containing $\Omega$, with the property that all the functions which are holomorphic on $\Omega$ are also holomorphic on $\Omega'$.} to the domain $\mathbb{R}^m+i\text{Ch}(A)$ where $\text{Ch}(A)$ is the smallest convex set containing the set $A$, called the convex hull of $A$.

By the above application, since $\bigcup_{\vec{\theta}}\mathcal{C}_\lambda^{\vec{\theta}}$ is non-convex we always have a holomorphic extension of $\bigcup_{\vec{\theta}}\mathcal{T}_\lambda^{\vec{\theta}}$ to a domain given by $\mathbb{R}^{(n-1)D}+i\text{Ch}(\bigcup_{\vec{\theta}}\mathcal{C}_\lambda^{\vec{\theta}})$. Furthermore $\mathcal{C}_\lambda$ contains $\text{Ch}(\bigcup_{\vec{\theta}}\mathcal{C}_\lambda^{\vec{\theta}})$ since $\mathcal{C}_\lambda$ is a convex cone containing $\bigcup_{\vec{\theta}}\mathcal{C}_\lambda^{\vec{\theta}}$. In subsequent subsections, we deal with the explicit cases of the three-, four- and five-point functions, where up to the four-point function we obtain that for each $\mathcal{C}_\lambda$ such extension yields the full of $\mathcal{C}_\lambda$, i.e. $\text{Ch}(\bigcup_{\vec{\theta}}\mathcal{C}_\lambda^{\vec{\theta}})=\mathcal{C}_\lambda$, and for five-point function we obtain subcases in which we are able to prove this equality.

For the two-point function (i.e. $n=2$), $\mathcal{T}_\lambda=\bigcup_{\vec{\theta}}\mathcal{T}_\lambda^{\vec{\theta}}$. That is, whenever $p\in\mathcal{T}_\lambda$, $\text{Im }p_1$ $(=-\text{Im }p_2)$ lies on some two dimensional Lorentzian plane.

\subsection*{Remarks}

The properties of the tube $\bigcup_{\vec{\theta}}\mathcal{T}_\lambda^{\vec{\theta}}$ as a domain in several complex variables have been used here to extend it holomorphically. From the work of \cite{LES2019}, we know that all the relevant Feynman diagrams (those which do not have any internal line of a massless particle) in the perturbative expansion of the $n$-point Green's function are analytic in the common tube $\bigcup_{\vec{\theta}}\mathcal{T}_\lambda^{\vec{\theta}}$. Hence our extension of $\bigcup_{\vec{\theta}}\mathcal{T}_\lambda^{\vec{\theta}}$ is valid to all orders in perturbation theory.

A proper application of Bochner's tube theorem requires us to thicken the connected tube $\bigcup_{\vec{\theta}}\mathcal{T}_\lambda^{\vec{\theta}}$ in order to make it open. But the thickened tubes (as in appendix \ref{thickunionTlamb}) are not identical for all the Feynman diagrams. However the intersection of all these thickened tubes corresponding to distinct diagrams certainly contains $\bigcup_{\vec{\theta}}\mathcal{T}_\lambda^{\vec{\theta}}$. We consider any relevant Feynman diagram. The corresponding thickened tube can be holomorphically extended to its convex hull due to Bochner's tube theorem. This extended domain contains the tube $\text{Ch}(\bigcup_{\vec{\theta}}\mathcal{T}_\lambda^{\vec{\theta}})=\mathbb{R}^{(n-1)D}+i\text{Ch}(\bigcup_{\vec{\theta}}\mathcal{C}_\lambda^{\vec{\theta}})$. Clearly, $\text{Ch}(\bigcup_{\vec{\theta}}\mathcal{T}_\lambda^{\vec{\theta}})$ lies in the intersection of all such extensions corresponding to distinct diagrams since $\text{Ch}(\bigcup_{\vec{\theta}}\mathcal{T}_\lambda^{\vec{\theta}})$ only includes convex combinations of points from $\bigcup_{\vec{\theta}}\mathcal{T}_\lambda^{\vec{\theta}}$. Hence $\text{Ch}(\bigcup_{\vec{\theta}}\mathcal{T}_\lambda^{\vec{\theta}})$ is the domain where all the relevant Feynman diagrams (at all orders in perturbation theory) are analytic.

\subsection{Three-point function}
\label{3ptfun}

For three-point function, we have $n=3$ and the sign-valued maps $\lambda(I)$ (described by equation (\ref{eq:lambprop})) can be given explicitly as follows. In this case, the primitive domain $\mathcal{D}$ essentially contains the union of 6 mutually disjoint tubes denoted by $\{\mathcal{T}_a^{(3)\pm},\ a=1,2,3\}$ and these primitive tubes are given by\footnote{Primitive tubes are generally defined in \eqref{eq:primTlamb}. Here, an additional superscript $^{(3)}$ in the notations for the tubes stands for the 3-point function.}
\begin{equation}
\label{eq:3ptTlamb}
  \mathcal{T}_a^{(3)\pm}=\bigg\{p\in\mathbb{C}^{2D}:\ \text{Im }p\in \mathcal{C}_a^{(3)\pm}\bigg\}\ ,
\end{equation}
where $p=(p_1,p_2,p_3)$ is linked by $p_1+p_2+p_3=0$. Their conical bases are defined by
\begin{equation}
\label{eq:3ptTbases}
  \mathcal{C}_a^{(3)+}=-\mathcal{C}_a^{(3)-}=\bigg\{\text{Im }p:\ \text{Im }p_b,\ \text{Im }p_c\in V^+\bigg\}\ ,
\end{equation}
where $(abc)=$ permutation of $(123)$. In order to define each of the above cones $\mathcal{C}_a^{(3)+}\ (\mathcal{C}_a^{(3)-})$, we require a certain pair of imaginary external momenta which (or their negative) are specified to be in the open forward lightcone $V^+$. For a given conical base, this in turn fixes all other $\text{Im }P_I$ to be in specific lightcone\footnote{For $n=3$, the total number of possible $P_I$ $=2^3-2=6$.}.

The cones (\ref{eq:3ptTbases}) reside on the manifold $\text{Im }p_1+\text{Im }p_2+\text{Im }p_3=0$. In order to assign coordinates to the points in $\mathcal{C}_a^{(3)+}$, let us choose $\{\text{Im }p_b,\ \text{Im }p_c\}$ as our set of basis vectors. On the other hand, for $\mathcal{C}_a^{(3)-}$, let us choose $\{-\text{Im }p_b,-\text{Im }p_c\}$ as our basis. With this, any of the above cones is contained in a $\mathbb{R}^{2D}$ and is of the following common form
\begin{equation}
  \mathcal{C}^{(3)}=\bigg\{\vec{Q}=(P_\alpha,P_\beta):\ P_\alpha,\ P_\beta\in V^+\bigg\}\ ,
\end{equation}
where any $\vec{Q}\in\mathcal{C}^{(3)}$ can be written as a $D\times2$ matrix given by
\begin{equation}
\label{eq:genpt3ptT}
  \vec{Q}\ =\ 
  \begin{pmatrix}
    P^0_{\alpha} & P^0_{\beta} \\
    P^1_{\alpha} & P^1_{\beta} \\
    \vdots & \vdots \\
    P^{D-1}_{\alpha} & P^{D-1}_{\beta}
  \end{pmatrix},
\end{equation}
with conditions $P^0_r>+\sqrt{\sum_{i=1}^{D-1}(P^i_r)^2}\quad \forall r=\alpha,\beta$ ensuring that both the columns belong to the forward lightcone $V^+$. Hence given a $\vec{Q}$, the quantities $P^0_r-\sqrt{\sum_i(P^i_r)^2},\ r=\alpha,\beta$ are a pair of positive numbers. Furthermore, the two columns of $\vec{Q}$ in general do not lie on a same two dimensional Lorentzian plane.

Now we consider cones $\mathcal{C}^{(3)\vec{\theta}}$ containing points $\vec{\tilde{Q}}$ where both the columns not only belong to $V^+$ but also lie on a same two dimensional Lorentzian plane $p^0-p^{\vec{\theta}}$ characterized by a $\vec{\theta}$. We shall show that taking points from these cones for various $\vec{\theta}$, a convex combination of them represents given $\vec{Q}$ in (\ref{eq:genpt3ptT}). This will complete the proof of $\text{Ch}(\bigcup_{\vec{\theta}}\mathcal{C}^{(3)\vec{\theta}})=\mathcal{C}^{(3)}$, since we already have $\mathcal{C}^{(3)}\supset\text{Ch}(\bigcup_{\vec{\theta}}\mathcal{C}^{(3)\vec{\theta}})$ as discussed right above the remarks in section \ref{HExtnLESD}.

We consider two points $\ \vec{\tilde{Q}}_1\in\mathcal{C}^{(3)\vec{\theta}_1}$ and $\ \vec{\tilde{Q}}_2\in\mathcal{C}^{(3)\vec{\theta}_2}$ given by
\begin{equation}
\label{eq:pts3ptTlamb}
  \vec{\tilde{Q}}_1\ =\ 
  2\begin{pmatrix}
    P^0_{\alpha}-\epsilon & \epsilon\\
    P^1_{\alpha} & 0\\
    \vdots & \vdots\\
    P^{D-1}_\alpha & 0\\
  \end{pmatrix},\quad
  \vec{\tilde{Q}}_2\ =\ 
  2\begin{pmatrix}
    \epsilon & P^0_{\beta}-\epsilon\\
    0 & P^1_{\beta}\\
    \vdots & \vdots\\
    0 & P^{D-1}_\beta\\
  \end{pmatrix},
\end{equation}
where we take any $\epsilon$ satisfying $0<\epsilon<\min\big\{P^0_r-\sqrt{\sum_{i=1}^{D-1}(P^i_r)^2},\ r=\alpha,\beta\big\}$ so that for each $r=\alpha,\beta$ we have $P^0_r-\epsilon>+\sqrt{\sum_{i=1}^{D-1}(P^i_r)^2}$. Both the columns of $\vec{\tilde{Q}}_1$ lie on a same two dimensional Lorentzian plane $p^0-p^{\vec{\theta}_1}$ where also the first column $P_\alpha$ of $\vec{Q}$ in (\ref{eq:genpt3ptT}) lies. Similarly, both the columns of $\vec{\tilde{Q}}_2$ lie on the two dimensional Lorentzian plane where $P_\beta$ lies. Now it is easy to check that the following relation holds
\begin{equation}
\label{eq:convcomb3ptfun}
  \frac{\vec{\tilde{Q}}_1}{2}+\frac{\vec{\tilde{Q}}_2}{2}=\vec{Q}\ .
\end{equation}
Equation (\ref{eq:convcomb3ptfun}) establishes that each of the 6 primitive tubes given in (\ref{eq:3ptTlamb}) can be obtained as holomorphic extension of a tube where the latter is contained in the LES domain $\tilde{\mathcal{D}}'$.

\subsection{Four-point function}
\label{4ptfun}

For four-point function, we have $n=4$ and the maps $\lambda(I)$ (described by equation (\ref{eq:lambprop})) can be given explicitly as follows. In this case, the primitive domain $\mathcal{D}$ essentially contains the union of 32 mutually disjoint tubes denoted by $\{\mathcal{T}_a^{(4)\pm},\ \mathcal{T}_{ab}^{(4)\pm},\ 1\leq a,b \leq4,\ a\neq b\}$\footnote{Here a superscript $^{(4)}$ in the notations for the tubes stands for the 4-point function.} and these primitive tubes are given by \cite{BEG1964,AB1960}
\begin{equation}
\label{eq:4ptTlamb}
\begin{aligned}
  \mathcal{T}_a^{(4)\pm}=\bigg\{p\in\mathbb{C}^{3D}:\ \text{Im }p\in \mathcal{C}_a^{(4)\pm}\bigg\},\quad \mathcal{T}_{ab}^{(4)\pm}=\bigg\{p\in\mathbb{C}^{3D}:\ \text{Im }p\in \mathcal{C}_{ab}^{(4)\pm}\bigg\}\ ,
\end{aligned}
\end{equation}
where $p=(p_1,\dots,p_4)$ is linked by $p_1+\cdots+p_4=0$. Their conical bases are defined by
\begin{equation}
\label{eq:4ptTbases}
\begin{aligned}
  &\mathcal{C}_a^{(4)+}=-\mathcal{C}_a^{(4)-}=\bigg\{\text{Im }p:\ \text{Im }p_b,\ \text{Im }p_c,\ \text{Im }p_d\in V^+\bigg\},\\
  &\mathcal{C}_{ab}^{(4)+}=-\mathcal{C}_{ab}^{(4)-}=\bigg\{\text{Im }p:\ -\text{Im }p_b,\ \text{Im }(p_b+p_c),\ \text{Im }(p_b+p_d)\in V^+\bigg\}\ ,
\end{aligned}
\end{equation}
where $(abcd)=$ permutation of $(1234)$. Note that in order to describe each of the above cones we require a certain set of three $\text{Im }P_I$ each of which (or its negative) is specified to be in the open forward lightcone $V^+$. For a given conical base, this in turn fixes all other $\text{Im }P_I$ to be in specific lightcone\footnote{For $n=4$, the total number of possible $P_I$ $=2^4-2=14$.} (see appendix \ref{4ptTbases}).

The cones (\ref{eq:4ptTbases}) reside on the manifold $\text{Im }p_1+\cdots+\text{Im }p_4=0$. Due to this link we can choose any three linear combinations of $\text{Im }p_1,\dots,\text{Im }p_4$ which are linearly independent as our set of basis vectors, to describe a given cone. In particular as our basis, we choose $\{\text{Im }p_b,\ \text{Im }p_c,\ \text{Im }p_d\}$ for the cones $\mathcal{C}_a^{(4)+}$, whereas we choose $\{-\text{Im }p_b,-\text{Im }p_c,-\text{Im }p_d\}$ for the cones $\mathcal{C}_a^{(4)-}$. Besides, as our basis, we choose $\{-\text{Im }p_b,\ \text{Im }(p_b+p_c),\ \text{Im }(p_b+p_d)\}$ for the cones $\mathcal{C}_{ab}^{(4)+}$\footnote{Instead, one can choose $\{\text{Im }p_b,\ \text{Im }p_c,\ \text{Im }p_d\}$ as the basis to describe points in any of the cones $\mathcal{C}_{ab}^{(4)+}$. This change of basis is a linear invertible transformation and the work of this subsection can be recast in this new basis (e.g., see appendix \ref{4ptTbases}).}, whereas we choose $\{\text{Im }p_b,-\text{Im }(p_b+p_c),-\text{Im }(p_b+p_d)\}$ for the cones $\mathcal{C}_{ab}^{(4)-}$. With this, any of the above cones is contained in a $\mathbb{R}^{3D}$ and is of the following common form
\begin{equation}
  \mathcal{C}^{(4)}=\bigg\{\vec{Q}=(P_\alpha,P_\beta,P_\gamma):\ P_\alpha,P_\beta,P_\gamma\in V^+\bigg\}\ ,
\end{equation}
where any $\vec{Q}\in\mathcal{C}^{(4)}$ can be written as a $D\times3$ matrix given by
\begin{equation}
\label{eq:genpt4ptT}
  \vec{Q}\ =\ 
  \begin{pmatrix}
    P^0_{\alpha} & P^0_{\beta} & P^0_{\gamma} \\
    P^1_{\alpha} & P^1_{\beta} & P^1_{\gamma} \\
    \vdots & \vdots & \vdots \\
    P^{D-1}_{\alpha} & P^{D-1}_{\beta} & P^{D-1}_{\gamma}
  \end{pmatrix},
\end{equation}
with conditions $P^0_r>+\sqrt{\sum_{i=1}^{D-1}(P^i_r)^2}\quad \forall r=\alpha,\beta,\gamma$ ensuring that each of the columns belong to the forward lightcone $V^+$. Hence given a $\vec{Q}$ the quantities $P^0_r-\sqrt{\sum_i(P^i_r)^2},\ r=\alpha,\beta,\gamma$ are three positive numbers. Furthermore the columns of $\vec{Q}$ in general do not lie on a same two dimensional Lorentzian plane.

Now we consider cones $\mathcal{C}^{(4)\vec{\theta}}$ containing points $\vec{\tilde{Q}}$ where all the three columns not only belong to $V^+$ but also lie on the same two dimensional Lorentzian plane $p^0-p^{\vec{\theta}}$ characterized by a $\vec{\theta}$. We shall show that taking points from these cones for various $\vec{\theta}$, a convex combination of them represents given $\vec{Q}$ in (\ref{eq:genpt4ptT}). This will complete the proof of $\text{Ch}(\bigcup_{\vec{\theta}}\mathcal{C}^{(4)\vec{\theta}})=\mathcal{C}^{(4)}$, since we already have $\mathcal{C}^{(4)}\supset\text{Ch}(\bigcup_{\vec{\theta}}\mathcal{C}^{(4)\vec{\theta}})$ as discussed right above the remarks in section \ref{HExtnLESD}.

We consider three points $\ \vec{\tilde{Q}}_1\in\mathcal{C}^{(4)\vec{\theta}_1}$, $\ \vec{\tilde{Q}}_2\in\mathcal{C}^{(4)\vec{\theta}_2}$ and $\ \vec{\tilde{Q}}_3\in\mathcal{C}^{(4)\vec{\theta}_3}\ $ given by
\begin{equation}
\label{eq:pts4ptTlamb}
\begin{aligned}
  \vec{\tilde{Q}}_1\ =&\ 
  3\begin{pmatrix}
    P^0_{\alpha}-\epsilon & \epsilon/2 & \epsilon/2\\
    P^1_{\alpha} & 0 & 0\\
    \vdots & \vdots & \vdots \\
    P^{D-1}_\alpha & 0 & 0\\
  \end{pmatrix},\quad
  \vec{\tilde{Q}}_2\ =\ 
  3\begin{pmatrix}
    \epsilon/2 & P^0_{\beta}-\epsilon & \epsilon/2\\
    0 & P^1_{\beta} & 0\\
    \vdots & \vdots & \vdots \\
    0 & P^{D-1}_\beta & 0\\
  \end{pmatrix},\\[5pt]
  &\quad \quad \quad \quad \ \vec{\tilde{Q}}_3\ =\ 
  3\begin{pmatrix}
    \epsilon/2 & \epsilon/2 & P^0_{\gamma}-\epsilon \\
    0 & 0 & P^1_{\gamma} \\
    \vdots & \vdots & \vdots \\
    0 & 0 & P^{D-1}_{\gamma} \\
  \end{pmatrix},
\end{aligned}
\end{equation}
where we take any $\epsilon$ satisfying $0<\epsilon<\min\big\{P^0_r-\sqrt{\sum_{i=1}^{D-1}(P^i_r)^2},\ r=\alpha,\beta,\gamma\big\}$ so that for each $r=\alpha,\beta,\gamma$ we have $P^0_r-\epsilon>+\sqrt{\sum_{i=1}^{D-1}(P^i_r)^2}$. All the three columns of $\vec{\tilde{Q}}_1$ lie on a same two dimensional Lorentzian plane $p^0-p^{\vec{\theta}_1}$ where also the first column $P_\alpha$ of $\vec{Q}$ in (\ref{eq:genpt4ptT}) lies. Similarly, all the columns of $\vec{\tilde{Q}}_2$ lie on the two dimensional Lorentzian plane where $P_\beta$ lies, and all the columns of $\vec{\tilde{Q}}_3$ lie on the two dimensional Lorentzian plane where $P_\gamma$ lies. Now it is easy to check that the following relation holds
\begin{equation}
\label{eq:convcomb4ptfun}
  \frac{\vec{\tilde{Q}}_1}{3}+\frac{\vec{\tilde{Q}}_2}{3}+\frac{\vec{\tilde{Q}}_3}{3}=\vec{Q}\ .
\end{equation}
Equation (\ref{eq:convcomb4ptfun}) establishes that each of the 32 primitive tubes given in (\ref{eq:4ptTlamb}) can be obtained as holomorphic extension of a tube where the latter is contained in the LES domain $\tilde{\mathcal{D}}'$.

\subsection{Five-point function}
\label{5ptfun}

For five-point function, we have $n=5$ and in this case the primitive domain $\mathcal{D}$ essentially contains the union of 370 mutually disjoint tubes whose conical bases are given by \cite{AB1960}\footnote{Here a superscript $^{(5)}$ in the notations for the cones stands for the 5-point function. And following \cite{AB1960} we use a prime only to distinguish between two classes of cones having same indices $(ab)$ or $(ab,c)$.}
\begin{equation}
\label{eq:5ptTbases}
\begin{aligned}
  &\mathcal{C}^{(5)+}_a=-\mathcal{C}^{(5)-}_a&=&\bigg\{\text{Im }p:\ \text{Im }p_b,\ \text{Im }p_c,\ \text{Im }p_d,\ \text{Im }p_e\in V^+\bigg\},\\
  &\mathcal{C}^{(5)+}_{ab}=-\mathcal{C}^{(5)-}_{ab}&=&\bigg\{\text{Im }p:\ -\text{Im }p_b,\ \text{Im }(p_b+p_c),\ \text{Im }(p_b+p_d),\ \text{Im }(p_b+p_e)\in V^+\bigg\},\\
  &\mathcal{C}'^{(5)+}_{ab}=-\mathcal{C}'^{(5)-}_{ab}&=&\bigg\{\text{Im }p:\ \text{Im }(p_a+p_c),\ \text{Im }(p_a+p_d),\ \text{Im }(p_a+p_e),\\
  &&&\quad\quad\quad\quad\quad\text{Im }(p_b+p_c),\ \text{Im }(p_b+p_d),\ \text{Im }(p_b+p_e)\in V^+\bigg\},\\
  &\mathcal{C}^{(5)+}_{ab,c}=-\mathcal{C}^{(5)-}_{ab,c}&=&\bigg\{\text{Im }p:\ \text{Im }p_c,\ -\text{Im }(p_b+p_c),\ \text{Im }(p_b+p_d),\ \text{Im }(p_b+p_e)\in V^+\bigg\},\\
  &\mathcal{C}'^{(5)+}_{ab,c}=-\mathcal{C}'^{(5)-}_{ab,c}&=&\bigg\{\text{Im }p:\ -\text{Im }(p_b+p_c),\ \text{Im }(p_a+p_c),\ \text{Im }(p_b+p_d),\ \text{Im }(p_b+p_e)\in V^+\bigg\},\\
  &\mathcal{C}^{(5)+}_{a,bc}=-\mathcal{C}^{(5)-}_{a,bc}&=&\bigg\{\text{Im }p:\ \text{Im }p_d,\ \text{Im }p_e,\ \text{Im }(p_a+p_b),\ \text{Im }(p_a+p_c)\in V^+\bigg\}\ ,
\end{aligned}
\end{equation}
where $(abcde)=$ permutation of $(12345)$ and $\text{Im }p=(\text{Im }p_1,\dots,\text{Im }p_5)$ is linked by the relation $\text{Im }p_1+\cdots+\text{Im }p_5=0$. Due to this link we can choose any four linear combinations of $\text{Im }p_1,\dots,\text{Im }p_5$ which are linearly independent as our set of basis vectors, to describe a cone which is given from the above list (\ref{eq:5ptTbases}). Hence any of these cones is contained in a $\mathbb{R}^{4D}$ with a choice for a basis.

We note that in order to describe each of the cones in (\ref{eq:5ptTbases}) except the cones $\mathcal{C}'^{(5)+}_{ab},\ \mathcal{C}'^{(5)-}_{ab}$, we require a certain set of four $\text{Im }P_I$ each of which (or its negative) is specified to be in the open forward lightcone $V^+$. This in turn fixes all other $\text{Im }P_I$ to be in specific lightcone\footnote{For $n=5$, the total number of possible $P_I$ $=2^5-2=30$.}. Now we confine ourselves to these cones which are 350 in numbers\footnote{Each of $\mathcal{C}'^{(5)+}_{ab}$ and $\mathcal{C}'^{(5)-}_{ab} $ is symmetric under the interchange of $a,b$ which is evident from (\ref{eq:5ptTbases}). Hence they are 20 in total.}. To describe any of these cones we choose the corresponding certain set of four $\text{Im }P_I$ as our basis (in a similar manner to the cases of the three-point and four-point functions, as demonstrated in detail in the sections \ref{3ptfun}, and \ref{4ptfun} respectively). With this, any of these cones is of the following common form
\begin{equation}
  \mathcal{C}^{(5)}=\bigg\{\vec{Q}=(P_\alpha,P_\beta,P_\gamma,P_\delta):\ P_\alpha,P_\beta,P_\gamma,P_\delta\in V^+\bigg\}\ ,
\end{equation}
where any $\vec{Q}\in\mathcal{C}^{(5)}$ can be written as a $D\times4$ matrix given by 
\begin{equation}
\label{eq:genpt5ptT}
  \vec{Q}\ =\ 
  \begin{pmatrix}
    P^0_{\alpha} & P^0_{\beta} & P^0_{\gamma} & P^0_{\delta} \\
    P^1_{\alpha} & P^1_{\beta} & P^1_{\gamma} & P^1_{\delta} \\
    \vdots & \vdots & \vdots & \vdots \\
    P^{D-1}_{\alpha} & P^{D-1}_{\beta} & P^{D-1}_{\gamma} & P^{D-1}_{\delta}
  \end{pmatrix},
\end{equation}
with conditions $P^0_r>+\sqrt{\sum_{i=1}^{D-1}(P^i_r)^2}\quad \forall r=\alpha,\beta,\gamma,\delta$. Given a $\vec{Q}$ as in (\ref{eq:genpt5ptT}) it can now be represented as the following convex combination.
\begin{equation}
\label{eq:convcomb5ptfun}
  \frac{\vec{\tilde{Q}}_1}{4}+\frac{\vec{\tilde{Q}}_2}{4}+\frac{\vec{\tilde{Q}}_3}{4}+\frac{\vec{\tilde{Q}}_4}{4}=\vec{Q}\ ,
\end{equation}
where $\vec{\tilde{Q}}_r,\ r=1,\dots,4$ are given by
\begin{equation}
\label{eq:pts5ptTlamb}
\begin{aligned}
  \vec{\tilde{Q}}_1\ =&\ 
  4\begin{pmatrix}
    P^0_{\alpha}-\epsilon & \epsilon/3 & \epsilon/3 & \epsilon/3 \\
    P^1_{\alpha} & 0 & 0 & 0 \\
    \vdots & \vdots & \vdots & \vdots \\
    P^{D-1}_\alpha & 0 & 0 & 0\\
  \end{pmatrix},\quad
  \vec{\tilde{Q}}_2\ =\ 
  4\begin{pmatrix}
    \epsilon/3 & P^0_{\beta}-\epsilon & \epsilon/3 & \epsilon/3 \\
    0 & P^1_{\beta} & 0 & 0 \\
    \vdots & \vdots & \vdots & \vdots \\
    0 & P^{D-1}_\beta & 0 & 0 \\
  \end{pmatrix},\\[5pt]
  \vec{\tilde{Q}}_3\ =&\ 
  4\begin{pmatrix}
    \epsilon/3 & \epsilon/3 & P^0_{\gamma}-\epsilon & \epsilon/3 \\
    0 & 0 & P^1_{\gamma} & 0 \\
    \vdots & \vdots & \vdots & \vdots \\
    0 & 0 & P^{D-1}_{\gamma} & 0 \\
  \end{pmatrix},\quad
  \vec{\tilde{Q}}_4\ =\ 
  4\begin{pmatrix}
    \epsilon/3 & \epsilon/3 & \epsilon/3 & P^0_{\delta}-\epsilon\\
    0 & 0 & 0 & P^1_{\delta}\\
    \vdots & \vdots & \vdots & \vdots \\
    0 & 0 & 0 & P^{D-1}_\delta \\
  \end{pmatrix},
\end{aligned}
\end{equation}
where we take any $\epsilon$ satisfying the condition: $0<\epsilon<\min\big\{P^0_r-\sqrt{\sum_{i=1}^{D-1}(P^i_r)^2},\ r=\alpha,\beta,\gamma,\delta\big\}$.

Equation (\ref{eq:convcomb5ptfun}) establishes that each of the primitive tubes described by (\ref{eq:5ptTbases}) except the ones whose conical bases are $\mathcal{C}'^{(5)+}_{ab},\ \mathcal{C}'^{(5)-}_{ab}$ can be obtained as holomorphic extension of a tube where the latter is contained in the LES domain $\tilde{\mathcal{D}}'$.

The above technique has limitations. Following difficulty arrises when we consider the remaining 20 cones which are given by $\mathcal{C}'^{(5)+}_{ab},\ \mathcal{C}'^{(5)-}_{ab}$. To describe points in $\mathcal{C}'^{(5)+}_{ab}$ let us choose the set
$$\bigg\{\text{Im }(p_a+p_c),\ \text{Im }(p_a+p_d),\ \text{Im }(p_a+p_e),\ \text{Im }(p_b+p_c)\bigg\}$$
as our basis, and to describe points in $\tilde{\mathcal{C}}'^{(5)-}_{ab}$ let us choose the set
$$\bigg\{-\text{Im }(p_a+p_c),\ -\text{Im }(p_a+p_d),\ -\text{Im }(p_a+p_e),\ -\text{Im }(p_b+p_c)\bigg\}$$
as our basis. With this any of the cones $\mathcal{C}'^{(5)+}_{ab},\ \mathcal{C}'^{(5)-}_{ab}$ is of the following common form
\begin{equation}
\label{eq:problemTbases}
 \mathcal{C}'^{(5)}=\bigg\{\vec{Q}=(P_\alpha,P_\beta,P_\gamma,P_\delta):\ P_\alpha,\ P_\beta,\ P_\gamma,\ P_\delta,\ (P_\beta+P_\delta-P_\alpha)\ \text{and}\ (P_\gamma+P_\delta-P_\alpha)\in V^+\bigg\}\ .
\end{equation}
Due to additional constraints on the linear combinations $(P_\beta+P_\delta-P_\alpha)$ and $(P_\gamma+P_\delta-P_\alpha)$ the technique which we have employed in earlier cases seems difficult to implement here analytically, in order to check the validity of $\text{Ch}(\bigcup_{\vec{\theta}}\mathcal{C}'^{(5)\vec{\theta}})=\mathcal{C}'^{(5)}$. Here each cone $\mathcal{C}'^{(5)\vec{\theta}}$ is to be obtained from $\mathcal{C}'^{(5)}$ by putting further restrictions on its points $\vec{Q}=(P_\alpha,P_\beta,P_\gamma,P_\delta)$ so that $\forall r=\alpha,\beta,\gamma,\delta\ \ P_r$ lies on the two dimensional Lorentzian plane $p^0-p^{\vec{\theta}}$. That is, it is difficult to find a set of points $\vec{\tilde{Q}}_r,\ r=1,\dots,m$ for some $m$\footnote{Here $m\leq 4D+1$, due to Carath\'eodory's theorem: if $A$ is a non-empty subset of $\mathbb{R}^q$, then any point of the convex hull of $A$ is representable as a convex combination of at most $q + 1$ points of $A$.}, each of which has four columns satisfying the six conditions as stated in (\ref{eq:problemTbases}) and furthermore all the four columns lie on a two dimensional Lorentzian plane, in such a way that a convex combination of these $m$ points produce a general point $\vec{Q}$ in (\ref{eq:problemTbases}). As an illustration we work with one of these 20 problematic cones in appendix \ref{problemTbases} (in which case, as a trial we take $m=4$).

\section{Limits within \texorpdfstring{$\mathcal{T}_\lambda$}{TEXT}}
\label{limTlamb}

In section \ref{HExtnLESD}, we have shown that for an $n$-point Green's function and given any $\lambda$ from the possible set $\Lambda^{(n)}$, the tube $\bigcup_{\vec{\theta}}\mathcal{T}_\lambda^{\vec{\theta}}$ has holomorphic extension inside the primitive tube $\mathcal{T}_\lambda$ where the former tube is contained in the LES domain $\tilde{\mathcal{D}}'$.

As per the equations (\ref{eq:LESD}) and (\ref{eq:complorentz}), if we take the limit $\text{Im }P_I\to0,\ \text{Im }P_I\in\mathcal{T}_\lambda^{\vec{\theta}}$ for a collection of subsets $\{I\}\subset\powerset^*(X)$, the $n$-point Green's function $G(p)$ in SFT is finite whenever we restrict their real parts by $-P_I^2<M_I^2$ for each $I$ belonging to that collection $\{I\}$. Here $\text{Re }P_J$ are kept arbitrary for all $J\in\powerset^*(X)\setminus\{I\}$. In fact, for a given collection $\{I\}$ by taking such limits within $\mathcal{T}_\lambda^{\vec{\theta}}$ for any $\vec{\theta}$ and restricting corresponding real parts, we reach to same value $G(p)$ for all $\vec{\theta}$. Now that $\bigcup_{\vec{\theta}}\mathcal{T}_\lambda^{\vec{\theta}}$ has an unique holomorphic extension given by $\text{Ch}(\bigcup_{\vec{\theta}}\mathcal{T}_\lambda^{\vec{\theta}})\subset\mathcal{T}_\lambda$, we reach to above value $G(p)$ in the limit $\text{Im }P_I\to0,\ \text{Im }P_I\in\text{Ch}(\bigcup_{\vec{\theta}}\mathcal{T}_\lambda^{\vec{\theta}})$ with above constraints on the real parts.

Evidently the family of holomorphic functions $\{G_\lambda(p),\ \lambda\in\Lambda^{(n)}\}$\footnote{Here $G_\lambda(p)$ denotes the analytic continuation of $G(p)$ defined on $\bigcup_{\vec{\theta}}\mathcal{T}_\lambda^{\vec{\theta}}$ to the domain $\text{Ch}(\bigcup_{\vec{\theta}}\mathcal{T}_\lambda^{\vec{\theta}})$.} defined on the family of mutually disjoint tubes $\{\text{Ch}(\bigcup_{\vec{\theta}}\mathcal{T}_\lambda^{\vec{\theta}}),\ \lambda\in\Lambda^{(n)}\}$ coincide on a real domain $\mathcal{R}$ given by
\begin{equation}
  \mathcal{R}=\bigg\{p\equiv(p_1,\dots,p_n):\ {\sum}_{a=1}^np_a=0\ \ \text{and}\ \ \forall I\in\powerset^*(X)\ -P_I^2<M_I^2\bigg\}\ .
\end{equation}

\section{Conclusions}

In this paper, we have shown that for any $n$-point Green's function in superstring field theory, the LES domain $\tilde{\mathcal{D}}'$ due to its shape always admits a holomorphic extension within the primitive domain $\mathcal{D}$ where the latter is basically the union of the convex primitive tubes. In the process we have found that the LES domain $\tilde{\mathcal{D}}'$ contains a non-convex connected tube within each convex primitive tube. The former tube being non-convex allows to include all the new points from its convex hull which is the set of all convex combinations of points in that tube. The convex tube thus obtained is a holomorphic extension of the former non-convex tube due to a classic theorem by Bochner, and lies inside the corresponding primitive tube.

Up to the four-point function such extension yields the full of the primitive domain. We have proved this result, in section \ref{3ptfun} for the three-point function obtaining all the 6 primitive tubes, and in section \ref{4ptfun} for the four-point function obtaining all the 32 primitive tubes. The appropriate real limits within those tubes in both cases are also attained (as discussed in section \ref{limTlamb}).

In section \ref{5ptfun}, we are able to show that for the five-point function such extension yields the full of 350 primitive tubes out of 370 primitive tubes which are possible in this case. The technique employed in this subsection can not be applied (as it is) for the remaining 20 primitive tubes, their shape being complicated. However within all these 370 extensions inside respective primitive tubes (obtaining 350 of them fully) the appropriate real limits are attained (as discussed in section \ref{limTlamb}).

As a consequence, our result shows that with respect to all the analyticity properties of the S-matrix which can be obtained relying on above extended domain inside the primitive domain, the infrared safe part of the S-matrix of superstring theory has similar behaviour to that of a standard local QFT. Any non-analyticity of the full S-matrix of SFT is entirely due to the presence of massless states --- which is also the case for a standard local QFT. The current approach is perturbative (because it uses Feynman diagrams) whereas the original proof of primitive analyticity for local QFTs is non-perturbative. Thus superstring amplitudes might also have potential singularities on the primitive domain arising from non-perturbative effects. Local QFTs are free from those. Furthermore, in local QFTs the following estimate holds on a primitive tube $\mathcal{T}_\lambda$ for each truncated cone $K^r\subset\mathcal{C}_\lambda\cup\{0\}$\footnote{A cone has been generally defined in footnote \ref{fn:cone}. Given a cone $K$ and a positive number $r$, a truncated cone is defined as (using $\parallel v\parallel$ to denote the Euclidean norm of $v$): $$K^r=\{v\in K:\ \parallel v\parallel\leq r\}\ .$$}.
\begin{equation}
\label{eq:estimateFL}
  \big\lvert G\left(\text{Re }p+i\text{Im }p\right)\big\rvert\ \leq\ A\frac{(1+\parallel\text{Re }p\parallel)^m}{\parallel\text{Im }p\parallel^l}\qquad\forall\ \text{Re }p\in\mathbb{R}^{(n-1)D},\ \text{Im }p\in K^r\setminus\{0\}\ ,
\end{equation}
where the numbers $A,m,l>0$ depend on $K^r$ \cite{Bogolyubov}\footnote{$\big\lvert z \big\rvert$ denotes the modulus of a complex number $z$.}. This is guaranteed as the off-shell $n$-point Green's function in a local QFT is equal to the Fourier-Laplace transform of some generalized function (more precisely, a tempered distribution) which is a position space correlator, and the analyticity of the off-shell Green's function on a primitive tube follows from causality constraints on the position space correlators. Equation \eqref{eq:estimateFL} is a place where superstring field theory could differ since typically its non-local vertices prevent us from defining position space correlators.

The difficulty arising for the twenty primitive tubes in the case for the five-point function has been demonstrated in appendix \ref{problemTbases}. This is a generic feature that arises for all higher-point functions, in the course of determining whether or not the application of Bochner's tube theorem yields certain primitive tubes fully. Solving this may require numerical analysis. We leave this for future work.

\section*{Acknowledgements}

We would like to thank Ashoke Sen for useful discussions. We also thank  Harold Erbin and Anshuman Maharana for useful comments on the draft.

\appendix

\section{Convexity of \texorpdfstring{$\mathcal{T}_\lambda,\ \mathcal{T}_\lambda^{\vec{\theta}}$}{TEXT}}
\label{convTlamb}

The tube $\mathcal{T}_\lambda$ (described by equation (\ref{eq:primTlamb})) is convex \cite{Pavlov1978}. We derive it as follows. We take any $m$ points $p^{(1)},\dots,p^{(m)}\in\mathcal{T}_\lambda$ and consider the convex combination $q=\sum_{r=1}^m t_r p^{(r)}$ where each $t_r\geq0$ and $\sum_{r=1}^m t_r=1$. Now if $q\in\mathcal{T}_\lambda$ then $\mathcal{T}_\lambda$ is convex.

Clearly $\sum_{a=1}^nq_a=0$. We define $Q_I=\sum_{a\in I}q_a$ for each $I\in\powerset^*(X)$. Hence we get $Q_I=\sum_{r=1}^m t_r P_I^{(r)}$ where for each $r$ we have $P_I^{(r)}=\sum_{a\in I} p^{(r)}_a$. We also have $\lambda(I)\text{Im }P^{(r)}_I\in V^+$ since $p^{(1)},\dots,p^{(m)}\in\mathcal{T}_\lambda$. Therefore $\lambda(I)\text{Im }Q_I=\sum_{r=1}^m t_r \lambda(I)\text{Im }P_I^{(r)}\in V^+$. Hence $q\in\mathcal{T}_\lambda$. This proves the convexity of $\mathcal{T}_\lambda$.

Taking any $m$ points $p^{(1)},\dots,p^{(m)}\in\mathcal{T}_\lambda^{\vec{\theta}}$ (described by equation (\ref{eq:LESTlamb})), now we show that the convex combination $q=\sum_{r=1}^m t_r p^{(r)}$ belongs to $\mathcal{T}_\lambda^{\vec{\theta}}$. In present case, since for each $r=1,\dots,m$ the $\text{Im }p^{(r)}_a\ \forall a$, in turn all $\text{Im }P_I^{(r)}$ lie on the two dimensional Lorentzian plane $p^0-p^{\vec{\theta}}$, therefore we get that the $\text{Im }Q_I$ for all $I$ including singletons lie on the same plane $p^0-p^{\vec{\theta}}$. Hence $\mathcal{T}_\lambda^{\vec{\theta}}$ is convex following similar steps to the above case.

\section{Nonconvexity of \texorpdfstring{$\bigcup_{\vec{\theta}}\mathcal{T}_\lambda^{\vec{\theta}}$}{TEXT}}
\label{nconvunionTlamb}

We take a point $p^{(1)}=(p^{(1)}_1,\dots,p^{(1)}_n)\in\mathcal{T}_\lambda^{\vec{\theta}=0}$. Hence $\text{Im }p^{(1)}_a\ \forall a=1,\dots,n$ lie on the two dimensional Lorentzian plane $p^0-p^1$. Now suppose the point $p^{(2)}$ is obtained by acting a real rotation\footnote{In equation \eqref{eq:actioncompLT}, we have defined actions of complex Lorentz transformations which include real rotations.} on the point $p^{(1)}$ so that $\text{Im }p^{(2)}_a\ \forall a=1,\dots,n$ now lie on the two dimensional Lorentzian plane $p^0-p^2$. Hence $p^{(2)}\in\mathcal{T}_\lambda^{\vec{\theta}_2}$ where $\vec{\theta}_2$ characterizes the two dimensional Lorentzian plane $p^0-p^2$. Now for the point $q=\frac{1}{2}p^{(1)}+\frac{1}{2}p^{(2)}$ which is the mid-point of the straight line segment connecting the two points $p^{(1)},\ p^{(2)}$ we have $\text{Im }q_a\ \forall a=1,\dots,n$ lying on the two dimensional Lorentzian plane $p^0-p^{\vec{\theta}_3}$ where the $p^{\vec{\theta}_3}$-axis lies on the two dimensional plane $p^1-p^2$ making an angle $45^0$ with the positive $p^2$-axis. Hence the point $q\in\mathcal{T}_\lambda^{\vec{\theta}_3}$. However we can change $\text{Im }p^{(2)}_1$ little bit, keeping all real parts and $\text{Im }p^{(2)}_a,\ a=2,\dots,n$ unchanged to obtain a new point $\tilde{p}^{(2)}$ such that $\tilde{p}^{(2)}$ still belongs to $\mathcal{T}_\lambda^{\vec{\theta}_2}$\footnote{In support of this see appendix \ref{thickunionTlamb}.}. Consequently we get two points $p^{(1)}\in\mathcal{T}_\lambda^{\vec{\theta}=0}$ and $\tilde{p}^{(2)}\in\mathcal{T}_\lambda^{\vec{\theta}_2}$, for which the mid-point of the straight line segment connecting them is given by $\tilde{q}=\frac{1}{2}p^{(1)}+\frac{1}{2}\tilde{p}^{(2)}$. Although $\text{Im }\tilde{q}_a\ \forall a=2,\dots,n$ lie on the two dimensional Lorentzian plane $p^0-p^{\vec{\theta}_3}$, the $D$-momenta $\text{Im }\tilde{q}_1=\frac{1}{2}p^{(1)}_1+\frac{1}{2}\tilde{p}^{(2)}_1$ does not lie on the two dimensional Lorentzian plane $p^0-p^{\vec{\theta}_3}$ anymore. Hence $\tilde{q}\notin\bigcup_{\vec{\theta}}\mathcal{T}_\lambda^{\vec{\theta}}$. In other words, $\bigcup_{\vec{\theta}}\mathcal{T}_\lambda^{\vec{\theta}}$ is non-convex.

To see this, let us take $\{p_1,\dots,p_{n-1}\}$ as our basis to describe points on the complex manifold $p_1+\cdots+p_n=0$. A generic point $p=(p_1,\dots,p_{n-1})$ on this manifold can be represented by an unique $D\times (n-1)$ matrix where the $a$-th column represent the $D$-momenta $p_a$. Since each $\mathcal{T}_\lambda^{\vec{\theta}}$ thereby $\bigcup_{\vec{\theta}}\mathcal{T}_\lambda^{\vec{\theta}}$ reside on this manifold now the imaginary parts of the points $p^{(1)}$, $p^{(2)}$, $\tilde{p}^{(2)}$, $q$ and $\tilde{q}$ can be represented in terms of $D\times (n-1)$ matrices as follows\footnote{For notational simplicity, we omit the prefix `$\text{Im}$' in all the entries in the equation \eqref{eq:nonconvex} however the entries should be understood as respective imaginary parts.}.
\begin{equation}
\label{eq:nonconvex}
\begin{aligned}
  &p^{(1)}=
  \begin{pmatrix}
    p^0_1 & p^0_2 & \cdots & p^0_{n-1} \\[5pt]
    p^1_1 & p^1_2 & \cdots & p^1_{n-1} \\[5pt]
    0 & 0 & \cdots & 0 \\[5pt]
    0 & 0 & \cdots & 0 \\
    \vdots & \vdots & \ddots & \vdots \\
    0 & 0 & \cdots & 0\\
  \end{pmatrix},\ 
  p^{(2)}=
  \begin{pmatrix}
    p^0_1 & p^0_2 & \cdots & p^0_{n-1} \\[5pt]
    0 & 0 & \cdots & 0 \\[5pt]
    p^1_1 & p^1_2 & \cdots & p^1_{n-1} \\[5pt]
    0 & 0 & \cdots & 0 \\
    \vdots & \vdots & \ddots & \vdots \\
    0 & 0 & \cdots & 0\\
  \end{pmatrix},\ 
  \tilde{p}^{(2)}=
  \begin{pmatrix}
    p^0_1 & p^0_2 & \cdots & p^0_{n-1} \\[5pt]
    0 & 0 & \cdots & 0 \\[5pt]
    \tilde{p}^1_1 & p^1_2 & \cdots & p^1_{n-1} \\[5pt]
    0 & 0 & \cdots & 0 \\
    \vdots & \vdots & \ddots & \vdots \\
    0 & 0 & \cdots & 0\\
  \end{pmatrix},\\[5pt]
  &\quad\quad\quad\quad\quad\quad\ q=
  \begin{pmatrix}
    p^0_1 & p^0_2 & \cdots & p^0_{n-1} \\[5pt]
    \frac{1}{2}p^1_1 & \frac{1}{2}p^1_2 & \cdots & \frac{1}{2}p^1_{n-1} \\[5pt]
    \frac{1}{2}p^1_1 & \frac{1}{2}p^1_2 & \cdots & \frac{1}{2}p^1_{n-1} \\[5pt]
    0 & 0 & \cdots & 0 \\
    \vdots & \vdots & \ddots & \vdots \\
    0 & 0 & \cdots & 0\\
  \end{pmatrix},\ 
  \tilde{q}=
  \begin{pmatrix}
    p^0_1 & p^0_2 & \cdots & p^0_{n-1} \\[5pt]
    \frac{1}{2}p^1_1 & \frac{1}{2}p^1_2 & \cdots & \frac{1}{2}p^1_{n-1} \\[5pt]
    \frac{1}{2}\tilde{p}^1_1 & \frac{1}{2}p^1_2 & \cdots & \frac{1}{2}p^1_{n-1} \\[5pt]
    0 & 0 & \cdots & 0 \\
    \vdots & \vdots & \ddots & \vdots \\
    0 & 0 & \cdots & 0\\
  \end{pmatrix}.
\end{aligned}
\end{equation}
Clearly all the columns of $\text{Im }q$ lie on the two dimensional Lorentzian plane $p^0-p^{\vec{\theta}_3}$ where the $p^{\vec{\theta}_3}$-axis lies on the two dimensional plane $p^1-p^2$ making an angle $45^0$ with the positive $p^2$-axis. It is also evident that the first column of $\text{Im }\tilde{q}$ does not lie on the two dimensional Lorentzian plane $p^0-p^{\vec{\theta}_3}$ although all the other columns of $\text{Im }\tilde{q}$ lie on it.

\section{Path-connectedness of \texorpdfstring{$\bigcup_{\vec{\theta}}\mathcal{T}_\lambda^{\vec{\theta}}$}{TEXT}}
\label{pconnecunionTlamb}

We take any two points $p^{(1)},\ p^{(2)}\in\bigcup_{\vec{\theta}}\mathcal{T}_\lambda^{\vec{\theta}}$. To show that the tube $\bigcup_{\vec{\theta}}\mathcal{T}_\lambda^{\vec{\theta}}$ is path-connected it is sufficient to find a path connecting the points $p^{(1)},\ p^{(2)}$ staying inside the tube $\bigcup_{\vec{\theta}}\mathcal{T}_\lambda^{\vec{\theta}}$.

Let $p^{(1)}\in\mathcal{T}_\lambda^{\vec{\theta}_1}$ and $p^{(2)}\in\mathcal{T}_\lambda^{\vec{\theta}_2}$ for some $\vec{\theta}_1,\vec{\theta}_2$. Now if $\vec{\theta}_1=\vec{\theta}_2\ (=\vec{\theta},\ \text{say})$ then $p^{(1)},\ p^{(2)}$ belong to the same tube $\mathcal{T}_\lambda^{\vec{\theta}}$. Since the tube $\mathcal{T}_\lambda^{\vec{\theta}}$ is convex (see appendix \ref{convTlamb}) the straight line segment connecting $p^{(1)},\ p^{(2)}$ lies entirely inside $\mathcal{T}_\lambda^{\vec{\theta}}$. Now we consider the case when $\vec{\theta}_1\neq\vec{\theta}_2$. In this case $p^{(1)}=(p^{(1)}_1,\dots,p^{(1)}_n)$, where all the $\text{Im }p^{(1)}_a$ thereby $\text{Im }P^{(1)}_I$ lie on the two dimensional Lorentzian plane $p^0-p^{\vec{\theta}_1}$. We consider the point $\tilde{p}^{(1)}=(\tilde{p}^{(1)}_1,\dots,\tilde{p}^{(1)}_n)\in\mathbb{C}^{(n-1)D}$ given by
\begin{equation}
\begin{aligned}
  \forall a\quad &\text{Im }\tilde{p}^{(1)0}_a=\text{Im }p^{(1)0}_a;\quad \text{Im }\tilde{p}^{(1)i}_a=0,\ i=1,\dots,D-1;\\
  &\text{Re }\tilde{p}^{(1)\mu}_a=\text{Re }p^{(1)\mu}_a,\ \mu=0,\dots,D-1\ ,
\end{aligned}
\end{equation}
where we switch off only the $\text{Im }p^{(1)i}_a,\ i=1,\dots,D-1$ components $\forall a$ in $p^{(1)}$ to obtain the point $\tilde{p}^{(1)}$. As the $\text{Im }p^{(1)0}_a$ component remains unaltered $\forall a$ the point $\tilde{p}^{(1)}\in\mathcal{T}_\lambda^{\vec{\theta}}$ for all $\vec{\theta}$. In particular $\tilde{p}^{(1)}$ belongs to both the tubes $\mathcal{T}_\lambda^{\vec{\theta}_1}$ and $\mathcal{T}_\lambda^{\vec{\theta}_2}$. Hence $\mathcal{T}_\lambda^{\vec{\theta}_1}$ being a convex tube the straight line segment $p^{(1)}\tilde{p}^{(1)}$ connecting $p^{(1)},\ \tilde{p}^{(1)}$ lies entirely inside $\mathcal{T}_\lambda^{\vec{\theta}_1}$, and $\mathcal{T}_\lambda^{\vec{\theta}_2}$ being a convex tube the straight line segment $\tilde{p}^{(1)}p^{(2)}$ connecting $\tilde{p}^{(1)},\ p^{(2)}$ lies entirely inside $\mathcal{T}_\lambda^{\vec{\theta}_2}$. Joining these two segments we get a path connecting $p^{(1)},\ p^{(2)}$ staying inside the tube $\bigcup_{\vec{\theta}}\mathcal{T}_\lambda^{\vec{\theta}}$. This completes the proof.

To see that $\tilde{p}^{(1)}\in\mathcal{T}_\lambda^{\vec{\theta}}$ for all $\vec{\theta}$, let us consider $\tilde{P}^{(1)}_I=\sum_{a\in I}\tilde{p}^{(1)}_a$ for an arbitrary non-empty proper subset $I$ of $X$. Therefore we have $\text{Im }\tilde{P}^{(1)}_I={\sum}_{a\in I}\text{Im }\tilde{p}^{(1)}_a$ and it is timelike because
\begin{equation}
\begin{aligned}
  \big(\text{Im }\tilde{P}^{(1)}_I\big)^2&=-\big({\sum}_{a\in I}\text{Im }\tilde{p}^{(1)0}_a\big)^2+{\sum}_{i=1}^{D-1}\big({\sum}_{a\in I}\text{Im }\tilde{p}^{(1)i}_a\big)^2\\
  &=-\big({\sum}_{a\in I}\text{Im }p^{(1)0}_a\big)^2<0\ .
\end{aligned}
\end{equation}
Furthermore $\text{Im }\tilde{P}^{(1)}_I$ and $\text{Im }P^{(1)}_I$ belong to the same lightcone because of the following
\begin{equation}
  \text{Im }\tilde{P}^{(1)0}_I={\sum}_{a\in I}\text{Im }\tilde{p}^{(1)0}_a={\sum}_{a\in I}\text{Im }p^{(1)0}_a=\text{Im }P^{(1)0}_I \implies \text{sgn}\big(\text{Im }\tilde{P}^{(1)0}_I\big)=\text{sgn}\big(\text{Im }P^{(1)0}_I\big)\ .
\end{equation}

\section{Thickening of \texorpdfstring{$\bigcup_{\vec{\theta}}\mathcal{T}_\lambda^{\vec{\theta}}$}{TEXT}}
\label{thickunionTlamb}

As mentioned in the introduction \ref{intro}, the work of \cite{LES2019} showed that at any point $p$ belonging to the LES domain $\mathcal{D}'$ all the relevant Feynman diagrams\footnote{As per the discussion in section \ref{intro}, diagrams that do not have any massless internal propagator are only relevant.} in the perturbative expansion of an $n$-point Green's function in SFT are analytic. Any such Feynman diagram at the point $p$ has an integral representation in terms of loop integrals where the poles of the integrand are at finite distance away from any of the loop integration contours. As per the discussion around equation (\ref{eq:complorentz}), the above statement also holds for any point $p$ belonging to the LES domain $\tilde{\mathcal{D}}'$.

Hence for any $p\in\bigcup_{\vec{\theta}}\mathcal{T}_\lambda^{\vec{\theta}}$ (thereby $\text{Im }p \in\bigcup_{\vec{\theta}}\mathcal{C}_\lambda^{\vec{\theta}}\subset\mathbb{R}^{(n-1)D}$) we can allow a small open ball $\mathcal{B}_{\text{Im }p}$ in $\mathbb{R}^{(n-1)D}$ centered at $\text{Im }p$ such that for any point $p'\in\mathcal{B}_{\text{Im }p}$ the aforementioned poles of the integrand are still at a finite distance away from the loop integration contours in a given Feynman diagram \cite{LES2019}. Consequently, the same integral representation in terms of loop integrals holds at any of the new points $p'$. Therefore, by allowing such open balls for each $p\in\bigcup_{\vec{\theta}}\mathcal{T}_\lambda^{\vec{\theta}}$ we can make $\bigcup_{\vec{\theta}}\mathcal{T}_\lambda^{\vec{\theta}}$ open, in which the given Feynman diagram still remains analytic. In this way, $\bigcup_{\vec{\theta}}\mathcal{T}_\lambda^{\vec{\theta}}$ can be thickened individually for all the relevant Feynman diagrams (at all orders in perturbation theory).

\section{The cone \texorpdfstring{$\mathcal{C}_{12}^{(4)+}$}{TEXT}}
\label{4ptTbases}

The cone $\mathcal{C}_{12}^{(4)+}$ taken from the list (\ref{eq:4ptTbases}) can be written as
\begin{equation}
  \mathcal{C}_{12}^{(4)+}=\bigg\{\text{Im }p:\ -\text{Im }p_2,\ \text{Im }(p_2+p_3),\ \text{Im }(p_2+p_4)\in V^+\bigg\}\ .
\end{equation}
$\mathcal{C}_{12}^{(4)+}$ resides on the manifold $\text{Im }p_1+\cdots+\text{Im }p_4=0$. Here we show how specifying $\text{Im }p_2$ in $V^-$\footnote{$V^-(=-V^+)$ is the open backward lightcone in $\mathbb{R}^D$.}, and $\text{Im }(p_2+p_3)$ and $\text{Im }(p_2+p_4)$ in $V^+$ in turn determine the sign-valued map $\lambda(I)$ (described by equation (\ref{eq:lambprop})) uniquely. Now we consider the following set of seven $\text{Im }P_I$
\begin{equation}
  \bigg\{\text{Im }p_2,\ \text{Im }p_3,\ \text{Im }p_4,\ \text{Im }(p_2+p_3),\ \text{Im }(p_2+p_4),\ \text{Im }(p_3+p_4),\ \text{Im }(p_2+p_3+p_4)\bigg\}\ .
\end{equation}
We want to see that for any $\text{Im }p$ inside the cone $\mathcal{C}_{12}^{(4)+}$ in which lightcone each element of the above set lies. Knowing this, similar information for any other possible $\text{Im }P_I$ can be determined using the relation $\text{Im }p_1+\cdots+\text{Im }p_4=0$. Now the following information can be obtained since for any $\text{Im }p\in\mathcal{C}_{12}^{(4)+}$ we have $-\text{Im }p_2,\ \text{Im }(p_2+p_3),\ \text{Im }(p_2+p_4)\in V^+$.
\begin{equation}
\label{eq:lamb4pt}
\begin{aligned}
  &\text{Im }p_3=-\text{Im }p_2+\text{Im }(p_2+p_3)\in V^+,\\
  &\text{Im }p_4=-\text{Im }p_2+\text{Im }(p_2+p_4)\in V^+,\\
  &\text{Im }(p_3+p_4)=\text{Im }p_3+\text{Im }p_4\in V^+,\\
  &\text{Im }(p_2+p_3+p_4)=-\text{Im }p_2+\text{Im }(p_2+p_3)+\text{Im }(p_2+p_4)\in V^+\ .
\end{aligned}
\end{equation}
Hence the signs $\lambda(I)$ corresponding to the cone $\mathcal{C}_{12}^{(4)+}$ is now known for any non-empty proper subset $I$ of $\{1,\dots,4\}$.

In section \ref{4ptfun}, with $\{-\text{Im }p_2,\ \text{Im }(p_2+p_3),\ \text{Im }(p_2+p_4)\}$ as our basis, points in the cone $\mathcal{C}_{12}^{(4)+}$ have been described. However any point in $\mathcal{C}_{12}^{(4)+}$ can uniquely be written in a new basis given by $\{\text{Im }p_2,\ \text{Im }p_3,\ \text{Im }p_4\}$ since such a change of basis is a linear transformation $\mathcal{L}$ with $\text{det}(\mathcal{L})=-1$. This transformation law can be stated as: any point $\vec{Q}=(P_\alpha,P_\beta,P_\gamma)$ written in the basis $\{-\text{Im }p_2,\ \text{Im }(p_2+p_3),\ \text{Im }(p_2+p_4)\}$ can be written as $\mathcal{L}\vec{Q}=(-P_\alpha,\ P_\alpha+P_\beta,\ P_\alpha+P_\beta)$ in basis $\{\text{Im }p_2,\ \text{Im }p_3,\ \text{Im }p_4\}$.

Hence the point $\vec{Q}$ in the cone $\mathcal{C}_{12}^{(4)+}$ as given in (\ref{eq:genpt4ptT}) in the new basis reads as
\begin{equation}
  \mathcal{L}\vec{Q}\ =\ 
  \begin{pmatrix}
    -P^0_{\alpha} & P^0_{\alpha}+P^0_{\beta} & P^0_{\alpha}+P^0_{\gamma} \\
    -P^1_{\alpha} & P^1_{\alpha}+P^1_{\beta} & P^1_{\alpha}+P^1_{\gamma} \\
    \vdots & \vdots & \vdots \\
    -P^{D-1}_{\alpha} & P^{D-1}_{\alpha}+P^{D-1}_{\beta} & P^{D-1}_{\alpha}+P^{D-1}_{\gamma}
  \end{pmatrix},
\end{equation}
with conditions $P^0_r>+\sqrt{\sum_{i=1}^{D-1}(P^i_r)^2}\quad \forall r=\alpha,\beta,\gamma$. And the points in (\ref{eq:pts4ptTlamb}) in the new basis read as
\begin{equation}
\begin{aligned}
  \mathcal{L}\vec{\tilde{Q}}_1\ =&\ 
  3\begin{pmatrix}
    -P^0_{\alpha}+\epsilon & P^0_{\alpha}-\epsilon/2 & P^0_{\alpha}-\epsilon/2\\
    -P^1_{\alpha} & P^1_{\alpha} & P^1_{\alpha}\\
    \vdots & \vdots & \vdots \\
    -P^{D-1}_\alpha & P^{D-1}_\alpha & P^{D-1}_\alpha\\
  \end{pmatrix},\quad
  \mathcal{L}\vec{\tilde{Q}}_2\ =\ 
  3\begin{pmatrix}
    -\epsilon/2 & P^0_{\beta}-\epsilon/2 & \epsilon\\
    0 & P^1_{\beta} & 0\\
    \vdots & \vdots & \vdots \\
    0 & P^{D-1}_\beta & 0\\
  \end{pmatrix},\\[5pt]
  &\quad \quad \quad \quad \quad \quad \quad \mathcal{L}\vec{\tilde{Q}}_3\ =\ 
  3\begin{pmatrix}
    -\epsilon/2 & \epsilon & P^0_{\gamma}-\epsilon/2 \\
    0 & 0 & P^1_{\gamma} \\
    \vdots & \vdots & \vdots \\
    0 & 0 & P^{D-1}_{\gamma} \\
  \end{pmatrix},
\end{aligned}
\end{equation}
where $\epsilon$ satisfies the condition: $0<\epsilon<\min\big\{P^0_r-\sqrt{\sum_{i=1}^{D-1}(P^i_r)^2},\ r=\alpha,\beta,\gamma\big\}$. Clearly above points $\mathcal{L}\vec{Q}$, $\mathcal{L}\vec{\tilde{Q}}_1$, $\mathcal{L}\vec{\tilde{Q}}_2$ and $\mathcal{L}\vec{\tilde{Q}}_3$ written in the basis $\{\text{Im }p_2,\ \text{Im }p_3,\ \text{Im }p_4\}$ are consistent with (\ref{eq:lamb4pt}). Furthermore each columns of $\mathcal{L}\vec{\tilde{Q}}_1$ lie on the same two dimensional Lorentzian plane where $P_\alpha$ lies. Similarly all the columns of $\mathcal{L}\vec{\tilde{Q}}_2$ lie on the two dimensional Lorentzian plane where $P_\beta$ lies, and all the columns of $\mathcal{L}\vec{\tilde{Q}}_3$ lie on the two dimensional Lorentzian plane where $P_\gamma$ lies. Now it is easy to check that the following relation holds
\begin{equation}
  \frac{\mathcal{L}\vec{\tilde{Q}}_1}{3}+\frac{\mathcal{L}\vec{\tilde{Q}}_2}{3}+\frac{\mathcal{L}\vec{\tilde{Q}}_3}{3}=\mathcal{L}\vec{Q}\ .
\end{equation}
Which depicts nothing but the linearity of $\mathcal{L}$ on (\ref{eq:convcomb4ptfun}).

\section{Difficulty arising for \texorpdfstring{$\mathcal{C}'^{(5)+}_{12}$}{TEXT}}
\label{problemTbases}

Consider a 5-point function, i.e. $n=5$. We take the following conditions which describe one of the 20 problematic cones, $\mathcal{C}'^{(5)+}_{12}$.
\begin{equation}
\label{eq:lamb5pt}
\begin{aligned}
  &\text{Im }(p_1+p_3),\ \text{Im }(p_1+p_4),\ \text{Im }(p_2+p_3),\ \text{Im }(p_2+p_4) \in V^+,\\
  &-\text{Im }(p_1+p_3+p_4),\ -\text{Im }(p_2+p_3+p_4)\in  V^+\ .
\end{aligned}
\end{equation}
Above conditions in turn imply that $\text{Im }p_1,\ \text{Im }p_2\in V^+$ and $\text{Im }p_3,\ \text{Im }p_4\in V^-$.

In this case, we choose $\big\{\text{Im }p_1,\ \text{Im }p_2,-\text{Im }p_3,-\text{Im }p_4\big\}$ as our basis to assign coordinates to the points in the cone $\mathcal{C}'^{(5)+}_{12}$. The goal is to find $P_1,P_2,P_3,P_4\in V^+$, in terms of which we can decompose a generic point $\big(\text{Im }p_1,\ \text{Im }p_2,-\text{Im }p_3,-\text{Im }p_4\big)$ in the cone in a convex combination of several points. Each of the terms in the decomposition should be in $\bigcup_{\vec{\theta}}\mathcal{C}'^{(5)+,\vec{\theta}}_{12}$.

Consider the following decomposition in a sum of four terms,
\begin{equation}
\label{eq:decomp}
\begin{aligned}
  &\big(\text{Im }p_1,\ \text{Im }p_2,-\text{Im }p_3,-\text{Im }p_4\big)=\ \big(\alpha_1 P_1,\ \alpha_2 P_1,\ \alpha_3 P_1,\ \alpha_4 P_1\big)\\
  &\quad\quad +\big(\beta_1 P_2,\ \beta_2 P_2,\ \beta_3 P_2,\ \beta_4 P_2\big)+\big(\gamma_1 P_3,\ \gamma_2 P_3,\ \gamma_3 P_3,\ \gamma_4 P_3\big)+\big(\delta_1 P_4,\ \delta_2 P_4,\ \delta_3 P_4,\ \delta_4 P_4\big)\ ,
\end{aligned}
\end{equation}
where $\alpha_r,\beta_r,\gamma_r,\delta_r,\ r=1,\dots,4$ all are real and non-negative. The conditions imposed by \eqref{eq:lamb5pt} imply for $\alpha_r$,
\begin{equation}
\label{eq:calpha}
  \alpha_1\geq\alpha_3,\ \alpha_4\ ;\quad \alpha_1\leq\alpha_3+\alpha_4\ ;\quad \alpha_2\geq\alpha_3,\ \alpha_4\ ;\quad \alpha_2\leq\alpha_3+\alpha_4\ .
\end{equation}
Exactly the same conditions should hold for $\beta_r,\gamma_r$ and $\delta_r$ as well. Note that in case of equality we can stay within the cone using $\epsilon$ prescription for the $p^0$ components. For example, suppose we have,
\begin{equation}
  \alpha_1=\alpha_3\ ;\quad \beta_1>\beta_3\ ;\quad \gamma_1>\gamma_3\ ;\quad \delta_1>\delta_3\ ,
\end{equation}
and all other inequalities (strictly) as in \eqref{eq:calpha}, without loss of generality. Then we can write the r.h.s. of \eqref{eq:decomp} as
\begin{equation}
\label{eq:epsprescrip}
\begin{aligned}
  &\big(\alpha_1 P_1+\tau\bar{\epsilon},\ \alpha_2 P_1,\ \alpha_3 P_1,\ \alpha_4 P_1\big)+\big(\beta_1 P_2-\tau\bar{\epsilon},\ \beta_2 P_2,\ \beta_3 P_2,\ \beta_4 P_2\big)\\
  &\quad\quad\quad +\big(\gamma_1 P_3,\ \gamma_2 P_3,\ \gamma_3 P_3,\ \gamma_4 P_3\big)+\big(\delta_1 P_4,\ \delta_2 P_4,\ \delta_3 P_4,\ \delta_4 P_4\big)\ ,
\end{aligned}
\end{equation}
where
\begin{gather}
  0<\tau<\min\big\{\beta_1-\beta_3,\ \beta_1-\beta_4\big\}\ ;\nonumber\\
  \bar{\epsilon}\equiv
  \begin{pmatrix}
    \epsilon\\
    0\\
    0\\
    \vdots\\
    0
  \end{pmatrix},
  \quad 0<\epsilon<\min\big\{P_r^0-|\vec{P}_r|,\ r=1,\dots,4\big\}\ .
\end{gather}
Clearly $P_2-\frac{\tau}{\beta_1}\bar{\epsilon},\ P_2-\frac{\tau}{\beta_1-\beta_3}\bar{\epsilon},\ P_2-\frac{\tau}{\beta_1-\beta_4}\bar{\epsilon}\in V^+$ since $\frac{\tau}{\beta_1},\ \frac{\tau}{\beta_1-\beta_3},\ \frac{\tau}{\beta_1-\beta_4}<1$. Hence for each term in \eqref{eq:epsprescrip}, we remain inside the cone $\tilde{\mathcal{C}}'^{+}_{12}$.

Now for each term in the decomposition \eqref{eq:decomp}, evidently all the four columns lie on a two dimensional Lorentzian plane. We need to solve for $P_1,P_2,P_3,P_4$ by inverting the following matrix equation,
\begin{equation}
\label{eq:matrixeq}
  \begin{pmatrix}
    \alpha_1 & \alpha_2 & \alpha_3 & \alpha_4 \\
    \beta_1 & \beta_2 & \beta_3 & \beta_4 \\
    \gamma_1 & \gamma_2 & \gamma_3 & \gamma_4 \\
    \delta_1 & \delta_2 & \delta_3 & \delta_4 
  \end{pmatrix}
  \begin{pmatrix}
    P_1\\
    P_2\\
    P_3\\
    P_4
  \end{pmatrix}=
  \begin{pmatrix}
    \text{Im }p_1\\
    \text{Im }p_2\\
    -\text{Im }p_3\\
    -\text{Im }p_4
  \end{pmatrix}.
\end{equation}
If we find that all the $P_r$ are in $V^+$, then the proof is done and we can say that $\mathcal{C}'^{(5)+}_{12}=\text{Ch}\big(\bigcup_{\vec{\theta}}\mathcal{C}'^{(5)+,\vec{\theta}}_{12}\big)$. But subject to conditions \eqref{eq:calpha}, solving \eqref{eq:matrixeq} seems to be difficult analytically.


\begin{thebibliography}{99}


\bibitem{LES2019}
  C.~de Lacroix, H.~Erbin and A.~Sen, ``Analyticity and crossing symmetry of superstring loop amplitudes,'' JHEP{\bf 05} (2019) 139 [arXiv:1810.07197].
  
\bibitem{LEKSV2017} 
  C.~de Lacroix, H.~Erbin, S.P.~Kashyap, A.~Sen and M.~Verma, ``Closed Superstring Field Theory and its Applications,'' Int. J. Mod. Phys. {\bf A 32} (2017) 1730021 [arXiv:1703.06410].
  
\bibitem{PS2016}
  R.~Pius and A.~Sen, ``Cutkosky rules for superstring field theory,'' JHEP {\bf 10} (2016) 024 [Erratum ibid. {\bf 1809} (2018) 122] [arXiv:1604.01783].
  
\bibitem{Bogolyubov}
  N.N.~Bogolyubov, A.A.~Logunov, A.I.~Oksak and I.T.~Todorov, ``General principles of quantum field theory,'' Mathematical physics and applied mathematics, Kluwer, Dordrecht The Netherlands (1990).
  
\bibitem{Steinmann1960}
  O.~Steinmann, ``\"Uber den Zusammenhang zwischen Wightmanfunktionen und retardierten Kommutatoren, I,'' Helv. Phys. Acta {\bf 33} (1960) 257.
  
\bibitem{Ruelle1961}
  D.~Ruelle, ``Connection between wightman functions and green functions in $p$-space,'' Nuovo Cim. {\bf 19} (1961) 356.
  
\bibitem{AB1960}
  H.~Araki and N.~Burgoyne, ``Properties of the momentum space analytic function,'' Nuovo Cim. {\bf 18} (1960) 342.
  
\bibitem{Araki1961}
  H.~Araki, ``Generalized Retarded Functions and Analytic Function in Momentum Space in Quantum Field Theory,'' J. Math. Phys. {\bf 2} (1961) 163.
  
\bibitem{BEG1964}
  J.~Bros, H.~Epstein and V.J.~Glaser, ``Some rigorous analyticity properties of the four-point function in momentum space,'' Nuovo Cim. {\bf 31} (1964) 1265.
  
\bibitem{BMS1961}
  J.~Bros, A.~Messiah and R.~Stora,`` A Problem of Analytic Completion Related to the Jost-Lehmann-Dyson Formula,'' J. Math. Phys. {\bf 2} (1961) 639.
  
\bibitem{BEG1965}
  J.~Bros, H.~Epstein and V.~Glaser, ``A proof of the crossing property for two-particle amplitudes in general quantum field theory,'' Commun. Math. Phys. {\bf 1} (1965) 240.
  
\bibitem{Bros1986}
  J.~Bros, ``Derivation of asymptotic crossing domains for multiparticle processes in axiomatic quantum field theory: a general approach and a complete proof for $2 \to 3$ particle processes,'' Phys. Rept. {\bf 134} (1986) 325.
  
\bibitem{Pavlov1978}
  V.P.~Pavlov, ``Analytic structure of the $3\to3$ forward amplitude,'' Theor. Math. Phys. {\bf 35}, 277–284 (1978). {\url{https://doi.org/10.1007/BF01032423}}.
  
\bibitem{MP1978}
  L.M.~Muzafarov and V.P.~Pavlov, ``Analyticity of the $3\to3$ forward amplitude in the neighborhood of the physical region,'' Theor. Math. Phys. {\bf 35}, 376–383 (1978). {\url{https://doi.org/10.1007/BF01039107}}.
  
\bibitem{LMMPS1979}
  A.A.~Logunov, B.V.~Medvedev, L.M.~Muzafarov, M.K.~Polivanov and A.D.~Sukhanov, ``Analytic structure of the $3\to3$ forward amplitude,'' Theor. Math. Phys. {\bf 40}, 677–687 (1979). {\url{https://doi.org/10.1007/BF01018717}}.
  
\bibitem{Bros1980}
  J.~Bros, ``Analytic structure of Green's functions in quantum field theory,'' in: Osterwalder K. (eds) Mathematical Problems in Theoretical Physics. Lecture Notes in Physics, vol 116. Springer, Berlin, Heidelberg (1980).
  
\bibitem{MPPS1982}
  B.V.~Medvedev, V.P.~Pavlov, M.K.~Polivanov and A.D.~Sukhanov, ``Analytic properties of many-particle amplitudes,'' Theor. Math. Phys. {\bf 52}, 723–732 (1982). {\url{https://doi.org/10.1007/BF01018410}}.
  
\bibitem{MPPS1984}
  B.V.~Medvedev, V.P.~Pavlov, M.K.~Polivanov and A.D.~Sukhanov, ``Analytic properties of multiparticle production amplitudes,'' Theor. Math. Phys. {\bf 59}, 427–440 (1984). {\url{https://doi.org/10.1007/BF01018176}}.
  
\bibitem{JL1957}
  R.~Jost and H.~Lehmann, ``Integral-Darstellung kausaler Kommutatoren,'' Nuovo Cim. {\bf 5} (1957) 1598.
  
\bibitem{Dyson1958}
  F.J.~Dyson, ``Integral representations of causal commutators,'' Phys. Rev. {\bf 110} (1958) 1460.
  
\bibitem{Bochner1937}
  S.~Bochner, ``A Theorem on Analytic Continuation of Functions in Several Variables,'' Annals of Mathematics, Second Series, {\bf 39} (1938), no. 1, 14--19. {\url{http://doi.org/10.2307/1968709}}.
  
\bibitem{Sen2017}
  A.~Sen, ``Wilsonian Effective Action of Superstring Theory,'' JHEP {\bf 01} (2017) 108 [arXiv:1609.00459].
  
\bibitem{Lassalle1974}
  M.~Lassalle, ``Analyticity properties implied by the many-particle structure of the $n$-point function in general quantum field theory. I. Convolution of $n$-point functions associated with a graph,'' Commun. Math. Phys. {\bf 36} (1974), no. 3, 185--226.
  
\bibitem{BL1975}
  J.~Bros and M.~Lassalle, ``Analyticity properties and many-particle structure in general quantum field theory. II. One-particle irreducible $n$-point functions,'' Commun. Math. Phys. {\bf 43} (1975), no. 3, 279--309.
  
\bibitem{Rudin1971}
  W.~Rudin, ``Lectures on the Edge-of-the-Wedge Theorem,'' CBMS Regional Conference Series in Mathematics, Volume 6, American Mathematical Society (1971).


\end{thebibliography}
\end{document}